\documentclass[Afour,sagev,times]{sagej}
\usepackage{moreverb,url}
\usepackage{enumitem}
\usepackage{stfloats}
\usepackage[colorlinks,bookmarksopen,bookmarksnumbered,citecolor=red,urlcolor=red]{hyperref}

\newcommand\BibTeX{{\rmfamily B\kern-.05em \textsc{i\kern-.025em b}\kern-.08em
T\kern-.1667em\lower.7ex\hbox{E}\kern-.125emX}}

\begin{document}

\runninghead{Darko and Park}

\title{Proactive Distributed Emergency Response with Heterogeneous Tasks Allocation}

\author{Justice Darko\affilnum{1} and Hyoshin Park\affilnum{2}}
\affiliation{
\affilnum{1}Information Technology Department, Rockwell Automation, Milwaukee, Wisconsin\\
\affilnum{2}Department of Engineering Management and Systems Engineering,\\ Old Dominion University, Norfolk, Virginia
}

\corrauth{Hyoshin Park, Ph.D, Associate Professor. Department of Engineering Management and Systems Engineering, Old Dominion University, Norfolk, Virginia}

\email{h1park@odu.edu}

\begin{abstract}
Traditionally, traffic incident management (TIM) programs coordinate the deployment of emergency resources to immediate incident requests without accommodating the interdependencies on incident evolutions in the environment. However, ignoring inherent interdependencies on the evolution of incidents in the environment while making current deployment decisions is shortsighted, and the resulting naive deployment strategy can significantly worsen the overall incident delay impact on the network. The interdependencies on incident evolution in the environment, including those between incident occurrences, and those between resource availability in near-future requests and the anticipated duration of the immediate incident request, should be considered through a look-ahead model when making current-stage deployment decisions. This study develops a new proactive framework based on the distributed constraint optimization problem (DCOP) to address the above limitations, overcoming conventional TIM models that cannot accommodate the dependencies in the TIM problem. Furthermore, the optimization objective is formulated to incorporate Unmanned Aerial Vehicles (UAVs). The UAVs' role in TIM includes exploring uncertain traffic conditions, detecting unexpected events, and augmenting information from roadway traffic sensors. Robustness analysis of our model for multiple TIM scenarios shows satisfactory performance using local search exploration heuristics. Overall, our model reports a significant reduction in total incident delay compared to conventional TIM models. With UAV support, we demonstrate a further decrease in the total incident delay ranging between 5\% and 45\% for the different number of incidents. UAV's active sensing can shorten response time of emergency vehicles, and a reduction in uncertainties associated with the estimated incident delay impact.
\end{abstract}

\keywords{Incident management, Emergency response, Distributed constraint optimization, Unmanned aerial vehicles, Look-ahead}

\maketitle
\section{Introduction}
Traffic incident management (TIM) programs have traditionally focused on reducing congestion and enhancing highway safety mainly through its Full-Function Service Patrols (FFSP)s program.\cite{fhwa_fsp} Under the FFSP program, service patrol vehicles such as Emergency Response Vehicles (ERV)s are dispatched to perform tasks including traffic incident clearance, traffic control and scene management, and incident detection and verification. In this study, we assume a trained FFSP operator uses a fully equipped ERV, capable of rapidly removing incident-involved automobiles or light trucks to a safe location without waiting for a wrecker.

One of the main challenges facing the TIM program is the efficient deployment of the limited resources (i.e., ERVs) in response to a sequence of incident requests. A typical deployment goal is to minimize the overall response time of all ERVs to incident locations and, therefore, minimize the travel time delay to commuters in the network. To achieve this goal, TIM programs have conventionally deployed the closest available ERV to an incident location and returned to the depot after completing the assignment.\citep{daneshgar2013evaluating}

However, this approach overlooks several key components that can improve the overall response time of ERVs to incident requests. First, to deploy an ERV to a current incident request, the dispatcher at the Traffic Operations Center (TOC) may consider interdependencies between past, current and possible near-future incident occurrences \citep{park2016real, pugh2019prediction} and allocate another ERV instead of the closest one to the current request. The closest available ERV not served in the current request could be closer to the location of the anticipated near-future emergency, possibly resulting in a lower overall response time of the ERVs to the sequence of incident requests and thus significantly decreasing the overall incident impact on the network.

Second, suppose a busy ERV is available earlier than the anticipated occurrence time of the next incident. Then, we can consider the ERV in the next deployment solution search. Conversely, if the anticipated availability time of a busy ERV is later than the occurrence time of the next incident, we exclude that ERV in the next deployment solution search. By considering the availability of the ERVs based on the anticipated incident duration, we avoid making shortsighted ERV assignments.

While considering the aforementioned components will improve the deployments of ERVs in TIM, emerging technologies such as Unmanned Aerial Vehicles (UAVs) are providing new opportunities to enhance operations in TIM. In particular, the Next-Generation (NextGen) TIM seeks to integrate new and emerging technology (e.g., UAVs), tools, and training to improve incident detection and safety response and reduce clearance times at roadway crashes.\citep{NEXTGEN_FHWA} Under revised Federal Aviation Authority (FAA) regulations that accommodate UAVs' advanced operations, UAVs and ERVs can coordinate their deployments in response to an incident request to provide more benefits in TIM programs. The role of UAVs in a heterogeneous vehicle team yields three key enhancements. First, aerial view of the possible route of ERVs to the incident location can update the availability status of the route and allow a safer and faster response to the incident location. Second, sparse freeway sensor networks cannot accurately estimate the location and speed of non-recurring congestion (e.g., shockwave).\citep{} Aerial monitoring of moving shockwaves can help estimate the true impact of an incident. Third, real-time information about an incident location can be gathered by monitoring the clearance progress from UAVs.   

Coordinating the deployment of a heterogeneous emergency vehicle team (i.e., ERVs and UAVs) to improve TIM is challenging. Previously, the distributed constraint optimization problem (DCOP), a generic modeling approach for multiagent systems and coordination, has been applied to problems such as the mobile sensor team, among other applications.\cite{Modi, Yeoh} This framework is an attractive problem-solving approach for coordinating the heterogeneous vehicle team in TIM for many reasons. 

First, the distributed constraint optimization framework is flexible in modeling the objectives of multiagent coordination problems through cost/utility constraints, overcoming the rigidity of models such as the constraint satisfaction model that defines a set of hard constraints and cannot easily model the dynamics in the TIM problem. The dynamics in TIM include varying numbers of available resources as they are deployed or busy serving other requests and the detection of new incidents when solving the current time problem. Second, the distributed framework for TIM can utilize the onboard computing units on the emergency vehicles and, with little communication overhead, \cite{} find an acceptable deployment scheme eliminating the need for a central computing station. 
Third, because the problem-solving activities are distributed among the agents (vehicles), the possibility of a single point of failure is eliminated.\citep{} Specifically, since knowledge of cost constraints are inherently distributed among the agents, each agent in the deployment fleet can autonomously decide the best solution by sharing minimum information. Lastly, the distributed framework can exploit potential parallelism in constraint networks that define the coordination between the two sub-teams of ERVs and UAVs in TIM. For example, by carefully connecting the problem of deploying ERVs and UAVs, the ERVs can use route-to-incident information from UAVs for a safe and fast response to incident scenes.

This paper develops the distributed constraint optimization framework for allocating ERVs and UAVs developed in a multi-agent TIM. We develop the distributed allocation of aerial and ground emergency resources by proactively responding to dynamic changes in the freeway network. The model's cost constraints directly consider the dispatch decision based on current incident requests in parallel with relocation of free ERVs based on the anticipated location of near-future emergencies. This unified approach will improve the overall benefit of deploying emergency resources by finding a solution that minimizes the sum of our developed cost constraints. The motivating domain is modeled as the Proactive Dynamic Routing Of uNmanned-aerial and Emergency Team Incident Management (P-DRONETIM). 
The main contribution of this paper and practical use in real-world scenarios are highlighted as follows:

\begin{itemize}[leftmargin=*]
	\item A distributed constraint decision-making framework for deploying emergency resources in TIM to minimize the delay impact of incidents on the network;
	
	\item A unified approach to dispatch ERVs based on current incident requests in parallel with relocation policies for free ERVs based on the anticipated location of near-future emergencies;
	
	\item The benefit of coordination between UAVs and ERVs is developed as a percentage reduction in response time of the ERV to the incident location based on the level of hazard on the ERV's route to the incident location.
	
	\item UAV aerial monitoring of moving shortwaves in the vicinity of an incident updates the estimate of the true impact of the incident by combining the estimates of the delay impact from the UAV observations and traditional road sensors in a data assimilation approach.
\end{itemize}

The best use of the proposed technology is ``Drone as First Responders'', which can station UAVs on the roof top of the building and launch with click of button to respond immediately upon receiving calls for service.  Video feed can provide what is going on so responders can make the better decisions. To the best of our knowledge, this is the first study that develops a proactive and dynamic model for the heterogeneous incident vehicle team in traffic incident management based on the DCOP approach. 

\color{black}

\section{Literature Review} 

\subsection{Overview of Optimization Technique}
We first provide an overview of the common state-of-the-art resource allocation algorithms from methodological perspective. These diverse approaches, from simulation frameworks to optimization algorithms, set the context for understanding the unique contributions of the proposed algorithms in later sections.

\subsubsection{Simulation-Based Framework \citep{deqi2012simulation}:}

The simulation frameworks are designed to emulate an emergency response system for highway traffic accidents with a primary focus on minimizing average response times across various accident scenarios. This involves the integration of dispatch strategies, allowing the simulation to account for dynamic traffic information, accident occurrence rates, and dynamic shortest path scenarios. In the pursuit of more realistic models, future efforts could be directed towards the development of models that better align with real-world characteristics. Additionally, exploring different effective algorithms for comparing performance and implementing the simulation framework using specific simulation tools are avenues for potential advancements.

\subsubsection{Bi-level Algorithms \citep{duan2015emergency,Yin, wang2020research}:}

These algorithms addresses the non-convex model for optimal fleet allocation in emergency responses. Employing a heuristic iterative approach, the model is treated as a master problem with a slave problem. The master problem involves minimizing a non-convex function with constraints on fleet allocation, and the slave problem is solved using a sensitivity-analysis-based iterative method. Despite lacking theoretical guarantees of convergence to a local optimum, empirical results showcase its effectiveness in practice. The algorithm's outer iteration utilizes a Sequential Quadratic Programming (SQP) algorithm, demonstrating its potential in optimizing fleet allocation for freeway service patrols.

\subsubsection{Scenario-Based Algorithms \citep{Yin1}:}

This algorithm introduces a decision framework tailored for optimal fleet allocation, particularly in addressing high-consequence incident response scenarios. Employing a risk-averse approach, it aims to minimize the expected loss associated with high-consequence scenarios. This is achieved by formulating the problem as a convex optimization problem, ensuring computational efficiency and enabling solutions in polynomial time. The loss function is defined based on the expected time for incident response, and regret associated with fleet allocation is determined by comparing its performance with optimal response times in each scenario. The algorithm strikes a balance between risk aversion and computational tractability.

\subsubsection{Work Function Algorithms with Look-ahead\citep{Park1}:}

This Algorithm enhances the performance of greedy algorithm in the allocation of emergency vehicles. The algorithm computes solutions incrementally using a dynamic programming approach, factoring in both total response time and a simple greedy decision process. An enhanced version, incorporates future information into the decision-making process. It aims to minimize a combination of the work function, response time, and expected future response time. This algorithm stands out by considering historical data and real-time updates, providing a comprehensive model for emergency vehicle allocation.

\subsection{State-of-the-art Traffic Incident Management}

\begin{table}[!ht]
	\centering
	\caption{The Proposed Approach to Overcome Challenges in Existing Traffic Incident Management}
	\vspace{-0.3em}
	\begin{tabular}{|c|c|c|}
		\hline
		$\text{Objectives}$ & $\text{Response only}$ &$\text{Total Delay}$  \\\hline
			Static    & Myopic\cite{deqi2012simulation, Yin,Yin1,wang2020research}  & Myopic\citep{JD2}    \\\hline
		Dynamic & Lookahead\citep{Park1}  & This study   	 \\\hline

	\end{tabular}
	\label{Tab:proposed}
\end{table}

Table \ref{Tab:proposed} summarizes the proposed dynamic approach for traffic incident management, highlighting its advantages over existing static methods in minimizing total delay. The Majority of research on highway incident management have focused on allocating a team of emergency vehicles by considering only the immediate incident request,\citep{deqi2012simulation, Yin,Yin1,wang2020research} without considering the interdependencies on the evolution of incidents in transportation networks. However, considering only the immediate request when deploying emergency vehicles is myopic and fails to accommodate the benefits of knowledge of the interdependencies on the evolution of incidents and its impact on improving the immediate request's decisions, thereby minimizing the overall incident impact on the network.

A few studies have attempted to address the above limitations. A notable example is the online optimization model that considers the stochastic nature of event occurrences.\citep{Park1} The model receives a sequence of emergency calls and performs a dispatch decision responding to each request. However, there is no direct framework for relocation of free emergency vehicles that can be made parallel with dispatch decisions. Other studies have developed the relocation, and dispatch decisions as separate models without considering the interdependencies between the two decisions.\citep{Park1,JD2} However, developing a connection between these two decisions will provide an efficient relocation decision to improve the overall response time during dispatch. The relocation of free emergency resources based on expected near-future emergencies can significantly improve the response time in TIM.

A key performance measure for TIM considering incident occurrences is the response time of emergency vehicles to the incident requests. Most studies have therefore proposed models that minimize the total response time for the team of emergency vehicles to all the sequence of incident requests. \cite{deqi2012simulation, Yin,Yin1,wang2020research,Park1,JD2} However, using just the response time as a performance measure is inadequate for assessing the true benefits of a deployment policy defined by the total delay impact on the network. Aside from the response time, the traffic characteristics at each incident location, including traffic flow rate, should be considered when making deployment decisions. In this study, we overcome the limitations of using the response time as the performance measure by directly modeling and utilizing the estimated total delay impact on the network as the performance measure.

To enhance the existing incident management programs, federal programs such as the NextGen TIM are integrating new and emerging technologies such as UAVs with the ground vehicles.\citep{NEXTGEN_FHWA} However, there has been a lack of focus on developing a framework to integrate UAVs to support the operations of ground vehicles in highway incident management. Under revised FAA regulations that accommodate UAVs' advanced operations,\citep{FAA_drones} a significant opportunity is presented for policy improvement in highway incident management. UAVs can provide enhanced information about the incident scene and aid the emergency vehicles in arriving at the incident location at full speed.

Previously, DCOP,\citep{Modi,Yeoh} a generic modeling approach for multiagent systems and coordination that has commonly been applied to problems such as the mobile sensor team, among others has been proposed for modeling the heterogeneous vehicle team in TIM.\citep{JD2} However, this model assumed a static environment, simplifying dynamic behaviors of traffic and events changing by time. A myopic resource allocation decision that does not consider the sequence of dependent events can not proactively adapt to changes in the environment over multiple time stages. Also, the availability of emergency resources for near-future emergencies was not considered that could be estimated based on the service time of the current request. The predictions of the environment and the resource availability in a future time should be considered through a look-ahead model when making decisions in response to the current request. 
\color{black}
Research on DCOP under stochastic uncertainty shows that network uncertainty can be reduced through multi-agent collaborative sampling.\citep{Leaute} Multi-agent coordination has been successfully used to efficiently explore unknown environments using the probabilistic multi-hypothesis tracker (PMHT) technique.\citep{Cheung} Using the Adopt polynomial-space algorithm, DCOP has been shown to achieve globally optimal solutions under asynchronous task execution while achieving high computational efficiency.\citep{Modi, Modi1} 

To accommodate the dynamically changing environment and the dependencies at different stages, this study model the resource allocation in the traffic incident management problem as the Proactive Dynamic Routing Of uNmanned-aerial and Emergency Team Incident Management (P-DRONETIM). P-DRONETIMs directly model interdependencies and the possible changes in the future stage when making current stage decisions. Compared to previous studies, we consider a heterogeneous vehicle fleet under a unified framework for dispatch and relocation decisions to accurately capture interdependencies in the decisions. The probability of incident occurrences at a location for different times of the day is achieved using incident characteristics of the given area. A look-ahead model considers the near-future availability of emergency vehicles and anticipates future events. 

DCOP solution approaches such as complete algorithms that find the optimal solutions have been proposed in literature.\citep{silaghi2006nogood,petcu2005dpop, modi2003asynchronous, maheswaran2004distributed,junges2008evaluating} However, complete algorithms are not practical for dynamic problems like TIM with large constraint density. On the other hand, incomplete algorithms such as the Maximum Gain Method (MGM) and the Distributed Stochastic Algorithm (DSA) provide high-quality solutions, especially for systems with large constraints density.\citep{Farinelli,grubshtein2010local,yin2009local} This study focuses on applying the incomplete algorithms MGM and DSA to P-DRONETIM.

In the event of natural disasters or sensor failures, UAVs serve as mobile sensors\cite{Parksensor22} supporting search and rescue efforts. For instance, UAVs establish a flying ad hoc network (FANET) to wirelessly communicate and cooperate. Numerous UAV applications are evolving to assist vehicular ad hoc networks (VANETs) and mobile ad hoc networks (MANETs)\cite{drones1,drones2,drones3}.

\section{Emergency Vehicle Team (EVT) in TIM}
We first describe the problem of the emergency vehicle team (EVT) in TIM. The EVT represented by the finite set of vehicles $\mathbb{A} = \{A_{1}, A_{2} , \dots, A_{n}\}$ are physically located in the transportation network. The network is modeled as a metric space on a grid with a function $d$ that defines the distance between two locations in the grid. The current location $cloc_{i}$ of each vehicle $A_{i}$ is assumed to be known. 

\begin{figure}[!htbp]
	\centering
	\includegraphics[width=0.94\columnwidth]{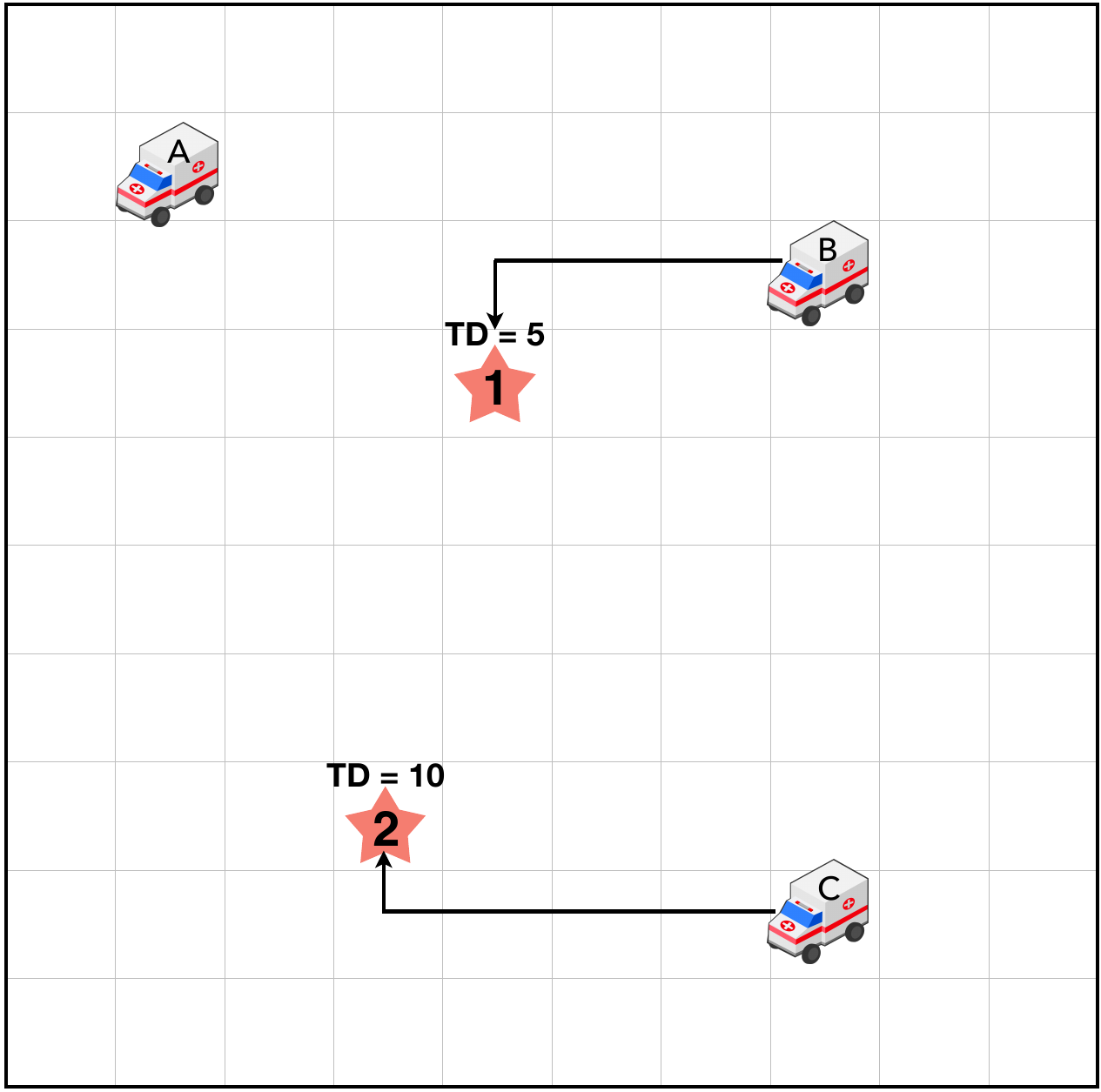}
	\caption{Emergency vehicle team in TIM with three agents and two incidents. Grid cells represent the possible locations an emergency vehicle can visit, and stars represent the incidents, characterized by their expected total delay on the network for each time step.}
	\vspace{-1.2em}
	\label{fig:evttim}	\vspace{0.7em}
\end{figure}

Considering the grid network, a finite set of locations defined by the grid cells represents the locations where a vehicle can assume a position or where an incident can occur. These locations represent a discretization of the overall underlying space on the transportation network. As shown in Figure \ref{fig:evttim} the cells in the grid network represent the possible locations for vehicles or incidents. We assume time is discretized so that the vehicles can estimate travel time between possible locations. 
The occurrence of an incident results in travel time delays for commuters at that location. We model the incident-induced delays on the network by a delay function $\mathrm{TD}$ that estimates the expected total delay impact on the network due to that incident. Indirectly we define incident locations $l$ as those whose $\mathrm{TD}_{l}$ is greater than zero. This definition can be strengthened by setting a threshold for $\mathrm{TD}_{l}$ since delays in the network can also be from non-incident events such as rush hours and social events. Because incidents may occur and be cleared by the EVT, $\mathrm{TD}_{l}$ changes dynamically as the EVT becomes aware of new incidents from the TOC. In the example presented in Figure \ref{fig:evttim} there are two incidents shown as star symbols, with estimated $\mathrm{TD}$ at the location. Given a set of incidents $l \in L$, the goal of the EVT is to be assigned in a way that minimizes the total delay impact of all the incidents in the network. In addition to the emergency vehicle response time to and clearance time of an incident, $\mathrm{TD}$ is a function of several other traffic parameters (e.g., traffic flow rate) defined in the vicinity of the incident location. We provide additional details of the delay function  $\mathrm{TD}$ in a later section. Depending on the number of available vehicles and incidents, the goal of the proactive and dynamic EVT in the TIM is to position themselves to minimize the total delay impact for all incidents considering the dynamics of vehicles and incidents and also predicting near-future incident occurrences.

\section{Proactive-DRONETIM Model}
\subsection{Approach overview}
Our model for the proactive and dynamic EVT in TIM is based on the DCOP framework. DCOP is defined by a tuple $\left\langle\mathbb{A}, \mathbb{X}, \mathbb{D}, \mathbb{F} \right\rangle$ where $\mathbb{A} = \left\{A_{1}, A_{2}, \ldots, A_{n}\right\}$ is a finite set of agents (emergency vehicles), $\mathbb{X} =\left\{X_{1}, X_{2}, \ldots ,X_{m}\right\}$ is a finite set of variables, $\mathbb{D} = \left\{D_{1}, D_{2}, \ldots, D_{m}\right\}$ is a finite set of domain values for each variable, $\mathbb{F}$ is a finite set of constraints between variables. Each variable $X_{i}$ is held by an agent who chooses a value to assign it from the finite set of values in domain $D_{i}$; Each constraint $\mathcal{C} \in \mathbb{F}$ is a function $C: D_{i_{1}} \times D_{i_{2}} \times \ldots \times D_{i_{k}} \rightarrow \mathbb{R}_{+} \cup\{0\}$ that maps assignments of a subset of the variables to a non-negative cost. The cost of a complete assignment of values to all variables is computed by aggregating the costs of all constraints. The agents have control over the assignment of values to their variables, and they are assumed to know only the constraints involving their variables, thus the distributed knowledge structure of DCOP. Agents coordinate by passing messages with other agents who hold variables constrained by their own variables called their neighbors. The goal of the DCOP is to find a sequence of assignments to variables with a minimum total cost. 

\subsection{Proactive Dynamic EVT in TIM as DCOP} 
The proactive and dynamic EVT in TIM problem is formulated as DCOP as follows; each vehicle holds a single variable for its position with a domain that includes all possible locations in the transportation network. 
A vehicle can be assigned to an incident in the current stage if it is available and not busy serving other incident locations. The total travel time required by the vehicle to reach the incident location and the clearance time for the incident determines the vehicles availability for a next assignment. Considering each incident location $l \in L$, a constraint $C_{l}$ relates the vehicles' current location to the delay impact $\mathrm{TD}_{l}$ of the incident in the network. The set of constraints $\mathbb{F}$ changes over time as incidents occur (TOC surveillance system discovers new incidents) and are cleared. In addition, the constraints cost can change over time due to changing impact of the incident on the network due to new and more accurate estimates of the impact from superior UAV surveillance systems. As discussed earlier, the $\mathrm{TD}$ function models the delay impact of incidents in the network. Realistically, the vehicles’ knowledge of the number of incidents and their impact on the network will be dynamic, reflecting the inherent dynamism in the TIM problem. We assume these inherent changes in the network are reported by the TOC. Vehicles that are available after the current stage assignment can be relocated to other locations in the network in anticipation of near-future changes in the $\mathrm{TD}$ function at those locations. We model the relocation decision by considering the dependencies in incident occurrences. Since we are only anticipating an incident at a given location, the benefits of such a decision are discounted to ensure that vehicles are first dispatched to the known incident locations before the relocation of the free vehicles, if any, in anticipation of a near-future incidents.

\subsubsection{Probability of incident occurrences.}
\begin{figure}[!htbp]
	\vspace{-1.0em}
	\centering
	\includegraphics[width=85mm]{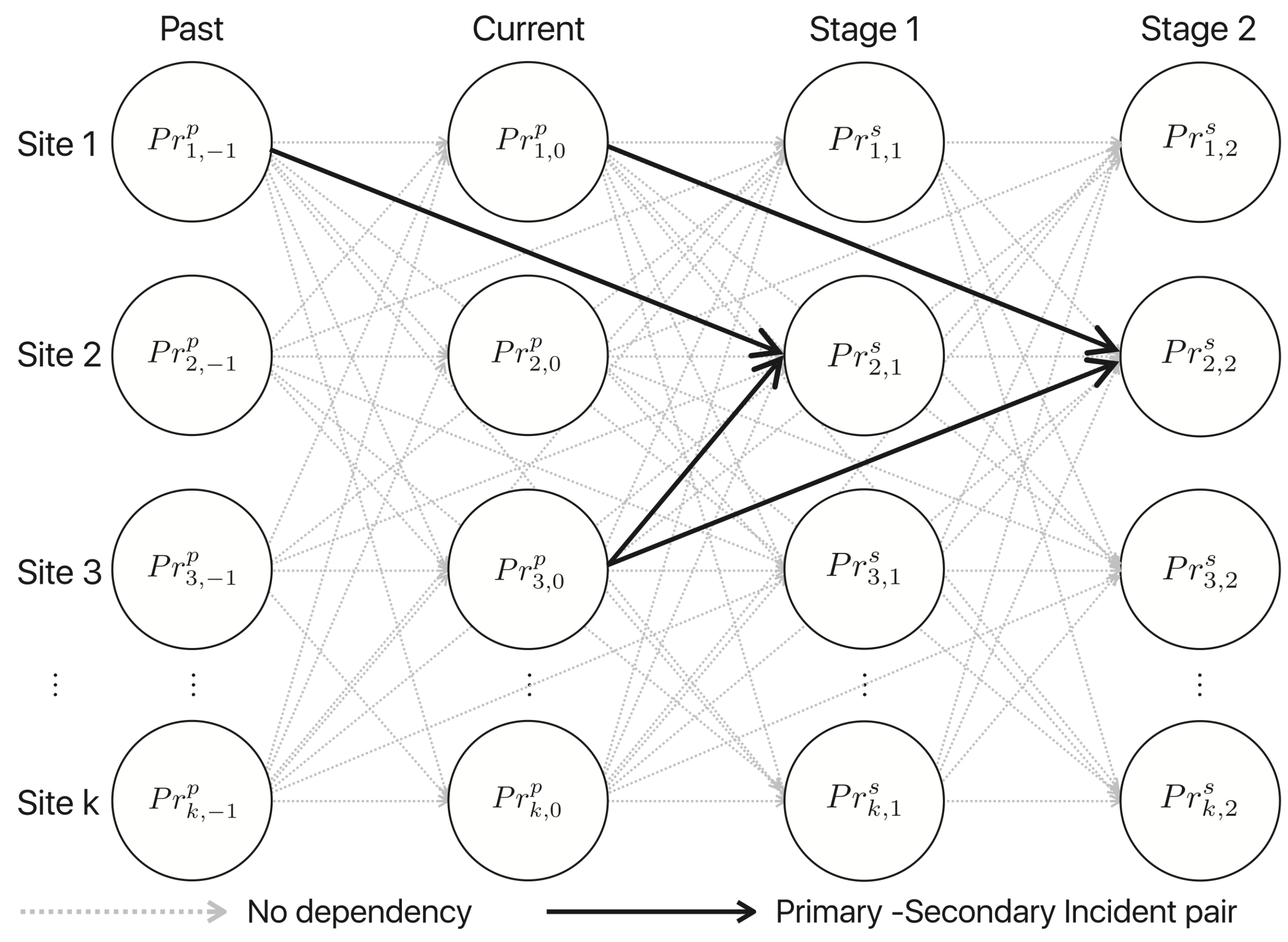}
	\caption{Dependency in incident occurrences.\citep{P2}}
	\label{fig:dependency}	\vspace{-0.8em}
\end{figure}

The relocation of free emergency vehicles are based on expected probability of near-future incident occurrences. 
For the set of possible locations $j \in J$, we define $\mathrm{Pr}_{j, r}$ as the probability of incident occurring at location $j$ in the next future stage $r$. 
The incident occurrence includes accumulated probabilities of secondary incidents in future stages, in which the impact of primary incidents overlaps. In general, a secondary incident may occur during the clearance or recovery of a primary incident. Therefore, we look-ahead two future stages. For example, the conditional probability of a secondary incident at site 2 at the first future-stage may depend on the probability of a primary incident at site 1 during the past and site 3 during the current stage; at the second-future stage may depend on the probability of a primary incident at site 1 and site 3 during the current stage (Figure \ref{fig:dependency}).

In this study the expected probability of incidents in the near-future stages is developed as presented in the previous study.\citep{P2} Let $\tau(j, r)$ be the normalized probability of incidents (probability of incidents at site $j$ over all locations $J$ in one stage) for each stage $r$. The expected probability of incidents $\mathbb{E}[\tau(j, r)]$ for each location $(j=k)$ and stage $(r=u)$ is a sum of $\mathrm{Pr}^{p}_{j,r}$ and $\mathrm{Pr}^{s}_{k,u}$. 
\begin{equation}
\mathbb{E}[\tau(j, r)]=\operatorname{Pr}_{j, r}^{p}+\operatorname{Pr}_{k, u}^{s} \quad \text { for } \quad j=k, r=u.
\vspace{-0.8em}
\end{equation}

$\mathrm{Pr}^{p}_{j,r}$ is the probability of primary and independent incidents at site $j$ during stage $r$, and $\mathrm{Pr}^{s}_{k,u}$ is the probability of secondary incident occurrences at location $k$ during stage $u$. The expected probability of incident at each location in the near-future stages is written as:
\begin{equation}
\begin{aligned}
\mathbb{E}[\tau(j,r)]=\operatorname{Pr}_{(j,r)}^{p}+\sum_{j} \delta(\Omega, \Delta)_{(j, r-1)(k,u)} \operatorname{Pr}_{j, r-1}^{p} \\
+\sum_{j} \delta(\Omega, \Delta)_{(j,r-2)(k,u)} \operatorname{Pr}_{j,r-2}^{p},
\end{aligned}
\label{eq:4}
\vspace{-0.8em}
\end{equation}

where the second and third terms is the explicit model for the probability of secondary incidents. The primary incident density ratio $\delta(\Omega, \Delta)_{(j, r)(k, u)}$ is defined to measure relative difference ratio and is not 0 only when an interrelation between incidents exists. P-DRONETIM estimates the expected probability of incident occurrence at all locations in the network in the first and second future stages using Equation \ref{eq:4}.

\subsubsection{Total delay impact on network due to incident.}\label{td}
The total delay, $\mathrm{TD}_{l}$, at incident location $l$ can be estimated using variables considered in the Highway Capacity Manual 2010.\citep{transportation2010trb} The deterministic incident delay model have underestimated the variance of actual delay and therefore the true impact of an incident in the freeway. In this study, the stochastic incident delay model is used to estimate the expected delay at incident location $l$ as:\citep{li2006estimation}\vspace{-1.3em}

\begin{equation}
\begin{split}
\mathbb{E}\left[\mathrm{TD}_{l}\left(t, r, s_{1}\right)\right]= 
\\
\frac{\left[\left(\bar{s}_{1}^{2}+\sigma_{s_{1}}^{2}\right)-(s+q) \cdot \bar{s}_{1}+s \cdot q\right] \cdot\left(\bar{r}^{2}+\sigma_{r}^{2}\right)}{2 \cdot(s-q)},
\end{split}
\label{eq:td}
\end{equation}

where the parameters in the model is defined as follows:

$s$ =  freeway capacity, which is also the departure rate after the
incident [vehicles per hour (vph)];

$\bar{s}_1$ = mean reduced freeway capacity during the incident (vph); 

$\sigma_{s_{1}}$ = standard deviation of reduced freeway capacity;

$q$ = traffic flow rate (vph);

$\bar{r}$ = mean incident duration and;

$\sigma_{r}$ = standard deviation of incident duration

The incident duration $r$ is the sum of the response time to and clearance time of the incident. The current location of the emergency vehicle is an important parameter for determining the response time and, therefore, the incident delay impact. Therefore considering multiple available vehicles, it is more reasonable to dispatch the closest available emergency vehicle to mitigate the delay impact. However, considering the global goal of minimizing the impact of all incidents, this may not always be the case. P-DRONETIM model considers the predictions of incident occurrences in future stages when making the current stage decisions. While this study assumes random values for clearance time for each incident based on historical crash data, more benefits will be achieved by accurately predicting the clearance time for the incident. Accurately predicting the clearance time for the incident improves the estimation of the incident delay impact on the network. Several approaches, including statistical and machine learning models, have predicted the expected clearance time of an incident which will be adopted for future study.\citep{tang2020statistical}
We assume that the above-listed traffic parameters are known for each incident location (e.g., from roadside sensors).

\subsection{TIM with UAV Active Sensing}
\begin{figure}[!htbp]
	\vspace{-0.5em}
	\centering
	\includegraphics[width=84mm]{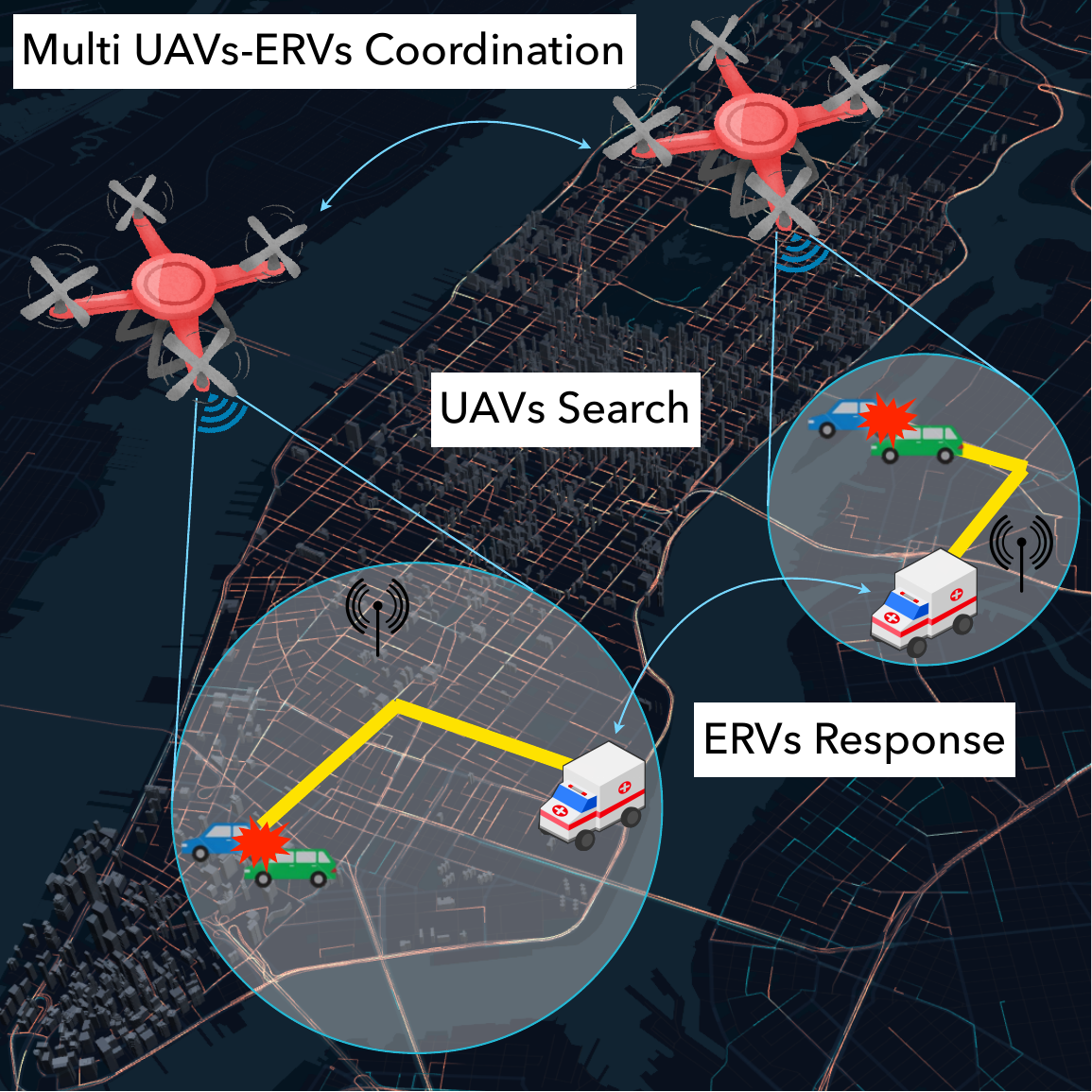}
	\caption{TIM with UAV active sensing.}
	\label{fig:uavsupport}
	\vspace{-0.9em}
\end{figure}

The role of a UAV sub-team in TIM is to explore unknown traffic conditions, unexpected situations on roads, and enhance prior information from loop detectors, automatic vehicle identification detectors, probe vehicle data, and other sensors (see Figure \ref{fig:uavsupport}). 

P-DRONETIM makes the following assumptions to ensure our model satisfies potential operational requirements and scenarios:
\begin{itemize}[leftmargin=*]
	\item UAVs have superior imaging technology (e.g., HD camera) and sensor payloads (e.g., infrared, LIDAR, etc.) and can perform high accuracy real-time surveillance by capturing video and images at higher elevations (within FAA regulations). In addition, the UAVs have a wider field of view (sensing range) and can provide enhanced information about an incident several miles ahead of ERV; 
	\item UAVs are launched from vehicles (e.g., trucks), serving as ground stations that have installed navigational computers and control stations for the UAVs. The trucks can be placed at strategic locations and moved when needed; 
	\item Depending on battery life, operators can bring UAVs back to the truck for battery exchange and relaunch; 
	\item The UAV ground stations have a two-way communication setup that allows communication with the TOC.
\end{itemize}  

The developed model deploys the UAVs by considering factors that enhance the operation of the EVT in TIM. Particularly to improve the response time to and clearance of incidents and assess the true impact (e.g., delays) of the incidents on the transportation network.

\subsubsection{Uncertainty of incident delay impact.}\label{uncertaintyreduction}

UAV deployment accounts for uncertainty in estimated incident delay impact. Drones, equipped with superior surveillance, observe real-time traffic queues and shockwaves, refining delay impact estimates. The model combines enhanced UAV information with prior delay estimates from traditional sensors, incorporating uncertainty defined by the variance equation $\sigma^{2}_{\mathrm{TD,P},l}$. Previous study have estimated this uncertainty as (parameters are defined as stated above):\citep{li2006estimation}\vspace{-1.8em}

\begin{equation}
\begin{split}
\sigma^{2}_{\mathrm{TD,P},l} = \operatorname{Var}\left[d\left(t, r, s_{1}\right)\right] = 
\frac{\left[\left(q-\bar{s}_{1}\right)^{2}+\sigma_{s_{1}}^{2}\right] \cdot\left(\sigma_{r}^{2}+\bar{r}^{2}\right)}{3 \cdot q^{2}}
\\-\frac{\left(q-\bar{s}_{1}\right)^{2} \cdot \bar{r}^{2}}{4 \cdot q^{2}}.
\end{split}
\end{equation}

Assuming that the variance of delay impact estimated from UAV observation is $\sigma^{2}_{\mathrm{TD,O},l}$, then the improvement by reducing uncertainty with enhanced UAV data is assessed using data assimilation. This technique adjusts variable estimates (e.g., delay impact) during the observation period \citep{asch2016data}. The combined estimate from traditional sensors and UAV observation yields the posterior uncertainty, represented as:

\begin{equation}
\sigma_{\text {TD, Posterior},l}^{2}=\left(1-\beta_{l} \right)\sigma^{2}_{\mathrm{TD,P},l}
\label{eq:be}
\end{equation}

The weight parameter $\beta$ considers the variance of the estimate from traditional sensors (prior), and the estimates from the UAV observation written as:
\begin{equation}
\beta_{l}=\frac{\sigma_{\mathrm{TD,P},l}^{2}}{\sigma_{\mathrm{TD,P},l}^{2}+\sigma_{\mathrm{TD,O},l}^{2}}
\end{equation}

Since the UAVs provide enhanced sensing capabilities than the traditional sensors, $\sigma_{\mathrm{TD, O},l}^{2}$ is expected to be less than $\sigma_{\mathrm{TD,P},l}^{2}$. Therefore, the posterior estimate of the uncertainty after UAV observation is always less than the prior as seen in Equation \ref{eq:be}. To update the expected delay impact (Equation \ref{eq:td}) after UAV observations, let $\mathrm{TD}_{\mathrm{P},l} = \mathbb{E}\left[\mathrm{TD}_{l}\left(t, r, s_{1}\right)\right]$ be the prior estimated total delay for incident location $l$ from traditional sensors, and $\mathrm{TD}_{\mathrm{O},l}$ estimated total delay from UAV observation, then the posterior estimate of the delay after UAV observation can be written as:\vspace{-0.6em}

\begin{equation}
\mathrm{TD}_{\text {Posterior},l}=\left(1-\beta_{l}\right) \mathrm{TD}_{\mathrm{P},l}+\beta_{l} \mathrm{TD}_{\mathrm{O},l}
\label{eq:be1}
\end{equation}

After UAV observations, the new information is transmitted to the TOCs, who can then update the TD function with the improved posterior estimate of the expected total delay impact from the incident.

\subsubsection{Status of ERVs route-to-incident.}\label{responetimeimp}

The TOC typically informs Emergency Response Vehicles (ERVs) about the best route to an incident. The UAV sub-team provides accurate information on this route, reducing ERVs' response time. The study assumes freeway shoulders as the primary route. Besides incident delay uncertainty, the P-DRONETIM UAV model considers how route information affects ERV response time and total delay reduction. With this support, ERVs can reach incidents faster, minimizing network delay impact. We make the following assumptions to justify the different levels of reduction in ERV response time when a UAV deploys to observe an ERV's route-to-incident location:

\begin{itemize}[leftmargin=*]
	\item The best recommended route-to-incident location has a hazard index (HI) based on past or current reported obstacles (e.g., disabled vehicles, debris, etc.) on the freeway shoulder. 
	
	\item The initial travel speed of the ERV is based on the hazard index for the route. For example, at a high hazard index, ERVs' are advised to travel at a slower speed to accommodate enough stopping distance in case of an obstacle and avoid a crash.
	
	\item An ERV will decelerate and change lane from shoulder to main travel lane and back to shoulder after passing the obstacle if there is one impeding the freeway shoulder. A lane change process briefly reduces the speed of ERV and thus the response time.
	
	\item A UAV deployed to monitor the ERVs' route-to-incident can confirm whether or not there is an obstacle on the freeway shoulder and the exact location of the obstacle if any.
\end{itemize}

This study assumes that the HI for a given route-to-incident location is categorized into five levels for simulation purposes. Each category has an assumed percentage reduction in response time with UAV support. 
$$
\mathrm{HI}=\left\{\begin{array}{ll}
{\text {1: very low }}             & \qquad{3\%} \\
{\text {2: low }}             &\qquad {5\%} \\
{\text {3: average }}             &\qquad {7\%} \\
{\text {4: high }}             &\qquad {9\%} \\
{\text {5: very high }} &\qquad {11\%}
\end{array}\right.
$$

For instance, if a UAV confirms a clear route for an Emergency Response Vehicle (ERV) with a Hazard Index (HI) of 5, the ERV can speed up, reducing response time by approximately 11\%. Effective coordination involves aligning UAV and incident locations, enabling ERVs to benefit from UAV information. Although each sub-team has its own task, they are both working towards a common goal

\subsubsection{Priority matrix for UAV deployment.}\label{prioritym}

A priority matrix is established to select optimal UAV deployment locations, incorporating incident severity, delay uncertainty, and hazard index, as outlined by previous studies \citep{P2}. The severity is associated with lane closures, and the model decomposes ERV and UAV tasks. Cooperation leads to reduced ERV response time when both sub-teams are assigned to the same location. 
The incident severity $\omega_{l}$ implicitly measures the impact of the incident on the transportation network. In this study, we define four levels of severity, where level one (low) indicates the lowest impact incident and level four the highest impact incident based on standard incident management definition.\citep{american2007manual, martin2011traffic}

$$
\omega_{l}=\left\{\begin{array}{ll}
{\text { low }}             & \qquad{1} \\
{\text { moderate }}      &\qquad {2} \\
{\text { high }}             &\qquad {3}\\
{\text { critical}}            &\qquad {4}
\end{array}\right.
$$

Finally, we simplify the development of the priority matrix by categorizing the prior uncertainty of the estimated delay based on the sparsity of the freeway sensor network $SS_{l}$ at the incident location $l$ which cannot accurately estimate the speed of non-recurring congestion. The higher the sparsity of the sensor network at the incident location, the less accurate the estimate of incident delay impact at the location. The levels are developed as ranging from 1 to 5, where level 1 indicates a very low uncertainty and level 5 indicates a very high uncertainty corresponding to low and high sparsity of sensor network at the incident location. 

$$
SS_{l}=\left\{\begin{array}{ll}
{\text { very low }}             & \qquad{1} \\
{\text { low }}             &\qquad {2} \\
{\text { average }}             &\qquad {3} \\
{\text { high }}             &\qquad {4} \\
{\text { very high }} &\qquad {5}
\end{array}\right.
$$

\begin{figure}[!htbp]
	\centering
	\includegraphics[width=84mm]{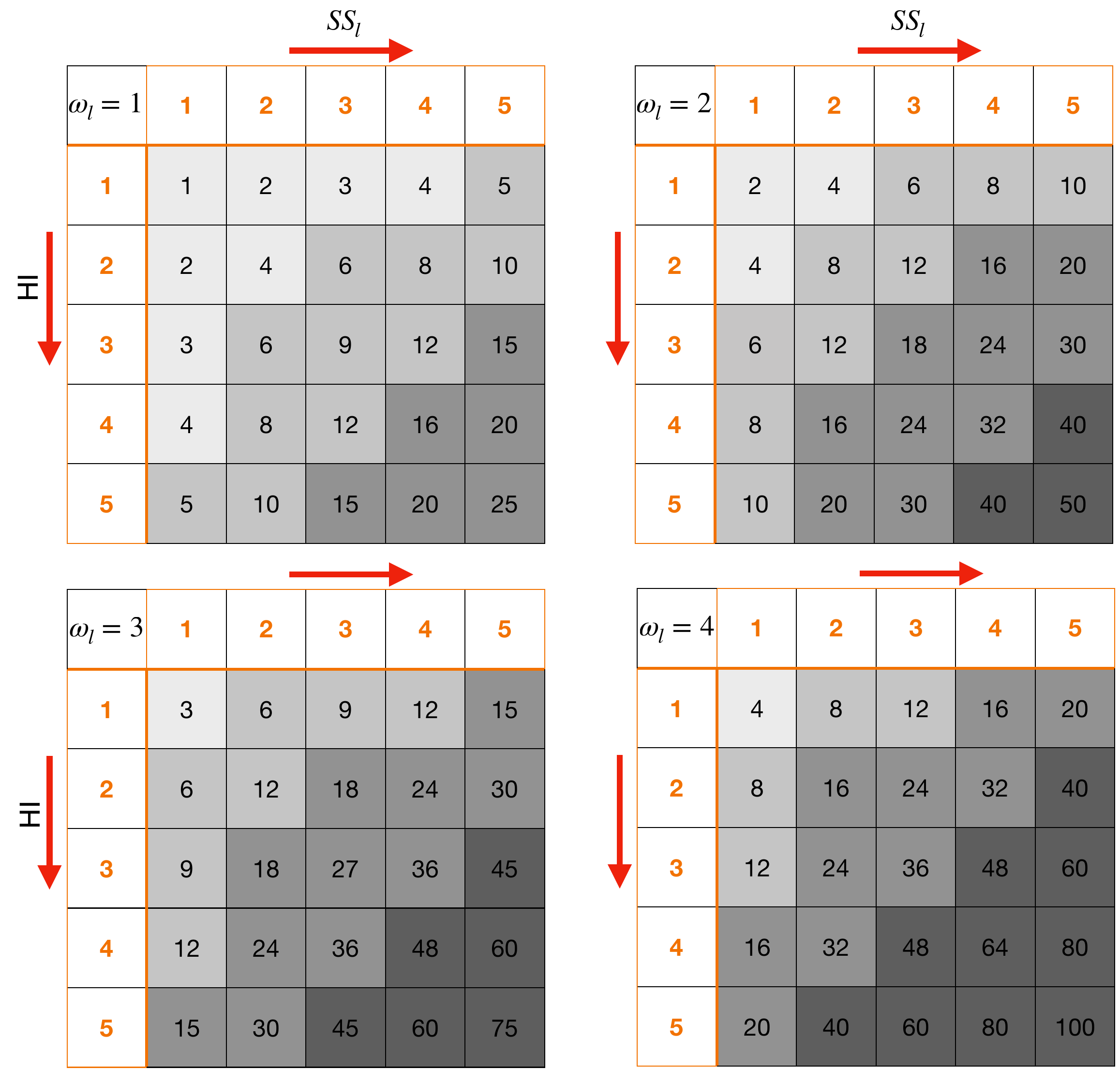}
	\vspace{-0.5em}
	\caption{UAV deployment priority matrix based on hazard index of the ERVs route-to-incident location $l$, delay impact uncertainty (sparsity of sensor network at incident location), and incident type (severity).}
	\vspace{-0.2em}
	\label{fig:pmatrix}
\end{figure}

P-DRONETIM quantifies the benefit for deploying a UAV to incident location $l$ as a mapping of the set of all ordered pair $\omega_{l} \times SS_{l} \times \mathrm{HI}$ to the real number $\mathbb{R}$. This approach is similar to the design of priority matrix for incident management that shows the importance of each task pair. In our case, the values indicate how much benefit the UAV will provide by deploying to a particular incident location (see Figure \ref{fig:pmatrix}). 

\subsection{Constraint relations}
\subsubsection{Ground emergency vehicle team.}\label{constraintEVT}
ERV deployment in P-DRONETIM aims to find the best solution that minimizes the sum of the defined constraint costs. Since each vehicle can communicate with all other vehicles in the network, the constraint graph is complete in structure. Specifically, each vehicle (node) is connected to every other vehicle in the EVT in the constraint graph. Each constraint is a function $C_{ij}: D_{i} \times D_{j} \rightarrow \mathbb{R}_{+} \cup\{0\} \text { (binary constraint) }$. The constraint relation between any two ERVs in P-DRONETIM is developed as:\vspace{-1.3em}

\begin{equation}
C\left(X_{a}, X_{b}\right)=\left\{\begin{array}{cc}
{ \infty } & {\text {if } X_{a} == X_{b}} \\
{ w_{j} c_{{X_{a}}}+ w_{j} c_{{X_{b}}}} & {\text {otherwise }}\\
\end{array}\right.
\label{eq13}
\end{equation}

where $X_{a}$ and $X_{b}$ are the variables held by ERVs $a$ and $b$. The weights $w_{j}$ is chosen from the set $\{ w_{d}, w_{r}\}$ dependent on the whether or not an incident has been reported at location $j$. Specifically, if the value assignment of variable $X_{i} = j$ and location $j$ has a reported incident ( i.e., $\mathrm{TD}_{j}>0$), the weight $w_{j}$ on the function $c_{X_{i}}$ is equal to $w_{d}$. The weight $w_{d}$ is applied to discount the cost for dispatch assignments. On the contrary, if location $j$ has no reported incident (i.e., $\mathrm{TD}_{j}\leq0$) then weight $w_{j}=w_{r}$ is applied to the function $c_{{X_{i}}}$. In this case, weight $w_{r} \texttt{>>} w_{d}$ is set to penalize such an assignment. That is,  increase the cost and indicate less importance. The values of the weight $w_{j}$ will ensure that relocation of emergency vehicles are carried out only after dispatch to reported incident locations are completed. 

Under limited resources, our developed constraint relations ensure that the emergency vehicles coordinate their local decision-making so that only one vehicle is assigned to a location at any time. Specifically, a partial or complete assignment of vehicles where $X_{a}$ equals $X_{b}$ incurs a very high cost ($\approx \infty$) to prevent or disincentive such assignment. For all other scenarios, two main criteria determine the cost for the constraint. Firstly, for $X_{i} = j$  where location $j$ has a reported incident, then $c_{{X_{i}}}$ is defined by Equation \ref{eq:td} (expected delay function). Secondly, if location $j$ has no reported incident, then $c_{{X_{i}}}$ is defined by considering the probability of an incident occurring at that location in the next time stage (Equation \ref{eq:4}). All locations with no reported incidents are candidates for relocation decisions for free emergency vehicles. In our formulation, because relocation decisions are designed to achieve an anticipated goal in a future stage, it is always less important than the goals of dispatching to true incidents in the current stage. 
The goal of PDRONETIM model is to find a sequence of assignments $\mathbf{X}^{*}$ for all the decision variables in $ \mathbb{X}$. Agents need to optimize a global objective function in a distributed fashion using local communication.\vspace{-1.0em}

\begin{equation}
\mathbf{X}^{*}=\underset{\mathbf{x} \in \Sigma}{\operatorname{argmin}} \quad \mathcal{C}^{h}(\mathbf{X})
\end{equation}

where $\Sigma$ is the assignment space for the decision variables of the P-DRONETIM.

\begin{equation}
\begin{aligned}
\mathcal{C}^{h}(\mathbf{X})=C_{a,b}\left(d_{a}, d_{b}\right)+\sum_{t=1}^{h}  C^{t}_{a,b}\left(d_{a}^{t}, d_{b}^{t}\right) \\
\quad X_{a} \leftarrow d_{a} \in D_{a}, X_{b} \leftarrow d_{b} \in D_{b}
\end{aligned}
\label{eq:13}
\end{equation}

Our model considers a look-ahead of two future stages ($h$=2). The first term $C_{a,b}\left(d_{a}, d_{b} \right)$ is the sum of the cost functions over the current stage. Looking at the reported incident set in the current time, our formulation of the cost allows the model to find the configuration of emergency vehicles for the incidents that will minimize the delay impact. The second term $\sum_{t=1}^{h} C^{t}_{a,b}\left(d_{a}^{t}, d_{b}^{t}\right)$ looks ahead into future stages  by evaluating the first and second future stage cost according to probability of incident incident occurring in that stage. Therefore, the configuration that minimizes the sum of the cost considering a look-ahead of the two future stages is the best deployment in the current stage.

\subsubsection{Unmanned aerial vehicle team.}
With the appropriate adjustment, the constraint relations developed for the EVT are extended for the UAVs. The goal is to find the best deployments for the UAVs to maximize their benefits for TIM. We define utility constraints that are based on the UAV deployment priority matrix  (Figure \ref{fig:pmatrix}). Similar to ERVs, the constraint relation for UAVs is developed as:\vspace{-1.3em}

\begin{equation}
U\left(X_{u}, X_{v}\right)=\left\{\begin{array}{cc}
{ -\infty } & {\text {if } X_{v} == X_{v}} \\
{ u_{{X_{u}}}+ u_{{X_{v}}}} & {\text {otherwise }}\\
\end{array}\right.
\label{eq:utilityUAV}
\end{equation}

where $X_{u}$ and $X_{v}$ are the variables held by UAVs $u$ and $v$. Since battery usage is a major limitation, UAV deployment is considered for only locations with reported incidents. Nonetheless, we can extend the model for surveillance to detect or monitor locations where incidents are anticipated by applying weights similar to those developed for the relocation of EVT. The utilities $u_{{X_{u}}}$ and $u_{{X_{v}}}$ for deployments of UAVs $u$ and $v$ are defined based on the priority matrix discounted by the distance of the  UAV to the incident location. Specifically, the travel distance of the UAV to incident location is subtracted from the computed benefit from the priority matrix.
The goal of the UAV deployment model is to find the optimal assignment to the incident locations as:
\begin{equation}
\begin{aligned}
\mathbf{U}^{*}=\operatorname{maximize} \sum U_{u,v} \left(X_{u}, X_{v}\right).
\end{aligned}
\label{eq:obj_uav}
\end{equation}

The domain $X_{u} \leftarrow d_{u} \in D_{u}, X_{v} \leftarrow d_{v} \in D_{v}$ for the variables held by the UAVs comprise of only locations of reported incidents. 
We have described two sub-teams performing tasks in the same network separately. However, it is reasonable to assume that cooperation between the sub-teams can lead to better results because although each sub-team has its task, they are both working towards a common goal of minimizing the delay impact of an incident in the network. For example, when an emergency vehicle has UAV support (assigned to the same incident), the estimated delay impact reduces since UAV support allows the ERV to reach the incident location safely and faster. In addition, the uncertainty of estimated delay impact improves (reduced), giving us a more accurate assessment of the impact of the incident on the network.

\section{SOLUTION APPROACH}
The local search algorithms MGM and DSA are used for solving P-DRONETIM, justified by the typical standards for choosing local over the complete search. Specifically time constraints and the limit on the scale of problems that complete algorithms can solve. Furthermore, P-DRONETIM also has unique characteristics that support using a local search algorithm. First, as required for a complete search, exploring the entire search space is not practical for large transportation networks with a large fleet of emergency resources. Second, the dynamic changes such as new incident requests while the current requests are being evaluated limit the time agents have to perform a complete algorithm because changes during the search for the best solution make the computed solution outdated.
In P-DRONETIM, the expected incident duration, given by the sum of the expected response and clearance time, is used to estimate the availability of the resource for future stage decisions. Therefore, not all possible assignment combinations will be feasible. The resource availability is directly considered for all sequences of value assignments for any variable. Specifically, a vehicle that is assigned multiple incidents in a sequence has already estimated its availability for the next assignment and, thus, how that impacts the total delay of the incidents in the network.  The main steps developed for the MGM and DSA solution approach are as follows:

\noindent \textbf{Step-1}: Make Initial random assignments for all vehicles. The vehicle initializes its information of the other vehicles' assignment in the network (initial is null).

\noindent \textbf{Step-2}: Each vehicle updates their information about the neighboring vehicles' assignment by receiving the current assignment of the neighbors. 

\noindent \textbf{Step-3}: Each vehicle computes its cost (estimated incident delay) for the current assignment while considering its information on the neighboring vehicles.

\noindent \textbf{Step-4}: Each vehicle then calculates its gain by finding the difference between the current assignment cost and the assignment that will result in the minimum cost.

\noindent \textbf{Step-5}: Each vehicle sends and receives the neighbor's gain. The vehicles then update the locations of the current assignments if their gain is greater than zero and the maximum gain of the neighbors. In the case of the DSA, the vehicle uses a stochastic strategy to decide whether or not it value assignment. Specifically, a probability threshold is set, and the vehicle's assignment is only updated if a randomly generated number is less than the threshold.

\noindent \textbf{Step-6}: Terminate if set number of iterations are performed else restart step 2 by sending a message that contains its new locations to its neighbors.

\begin{table*}[!tbh]\small
	\centering
	\caption{Simulation Parameters}
	\vspace{-0.3em}
	\begin{tabular}{|c|c|c|c|c|c|c|}
		\hline
		$\text{severity}$ & $s$ (vph)&$\bar{s}_{1}$(vph) & $\sigma_{s_{1}}$ (vph) &$q$ (vph)&$\sigma^{2}_{r}$ (h)&$\text{clearance time}$ (h)\\\hline
		1    & [750,800]& [600,800] &[100,200] &[600,720]&[0.1,0.2] &[0.2,0.3]	 \\\hline
		2    & [1130,1500] & [900,1900] &[100,300] &[960,1120] &[0.2,0.3]&[0.3,0.4]	 \\\hline
		3    & [1700,1900]&[1000,1200] &[100,300] &[1440,1644] &[0.2,0.4]&[0.5,0.7]	 \\\hline
		4    & [2200,2800] & [1000,1500] & [100,300] &[1824,2015] &[0.2,0.3]&[0.5,1]	 \\\hline
	\end{tabular}
	\label{Tab:parameters}
\end{table*}

\section{SCENARIOS AND EXPERIMENTS}
The proposed P-DRONETIM model is evaluated in a prototype network abstracted with an existing highway network's traffic parameter data. The problem scenarios are simulated in a 10-by-10 city block-like grid network, with 180 links representing the road segments and 100 points of intersections representing the possible crash locations. The weight on each edge, representing the average time to traverse the edge, is randomly drawn from a uniform distribution between 0.1 and 1.5 (hr). The total delay for an incident request is estimated based on Equation \ref{eq:td} with parameters described in Table \ref{Tab:parameters} and the response time to the incident estimated from a distance function on the grid network. Each incident is given a random severity number drawn from a uniform distribution between 1 and 4, which determines the values for traffic parameters used to estimate the total delay of the incident. For example, looking at Table \ref{Tab:parameters}, an incident request of severity one will have parameters drawn uniformly from the ranges defined for each of the variables in row one. The incidents are randomly located in the network through a random draw from a uniform distribution whose range is the size of the grid network. We assume each ERV and UAV is given an initial random location following current practice. The expected probability of an incident at different locations in the network is estimated based on Equation \ref{eq:4}. We randomly generate probabilities of incident occurrence for the near future stage for simulation purposes dependent on each current stage request. The incident interdependency indicator between locations and stages is randomly drawn between 0 and 1 for the pairs of locations and stages. A highly functional UAV that supports reliable communication over long distances with the TOCs is assumed for this study. 

A typical example is the Autel DragonFish Fixed Wing UAV, featuring Beyond Visual Line of Sight (BVLOS) UAV, Vertical Takeoff and Landing (VTOL), 120 Minute Flight Time, 18.6 Mile Transmission Range, and 50x Optical Zoom Capability. For simplicity, we assume that UAVs maintain a fixed altitude during their flights and cover a given region in the network once assigned to a given location. Vehicles can communicate with each other using a low-bandwidth Radio frequency. 
\color{black}

\section{RESULTS AND DISCUSSIONS}
The results for different TIM scenarios are first presented for the MGM and DSA approach to P-DRONETIM to assess its performance. The MGM algorithm is highly exploitative, preventing the exploration of other possible solutions. On the contrary, DSA uses a stochastic exploration technique to find other solutions that might be optimal.\citep{tel} In DSA, a vehicle's assignment to a location is updated only if a specified probability threshold condition has been satisfied, introducing randomness in selecting other feasible solutions. In our assessments, each scenario is repeated in ten experiments, and each solution point represents the mean and standard error values for that scenario.

\begin{figure}[!htbp]
	\centering
	\includegraphics[width=85mm]{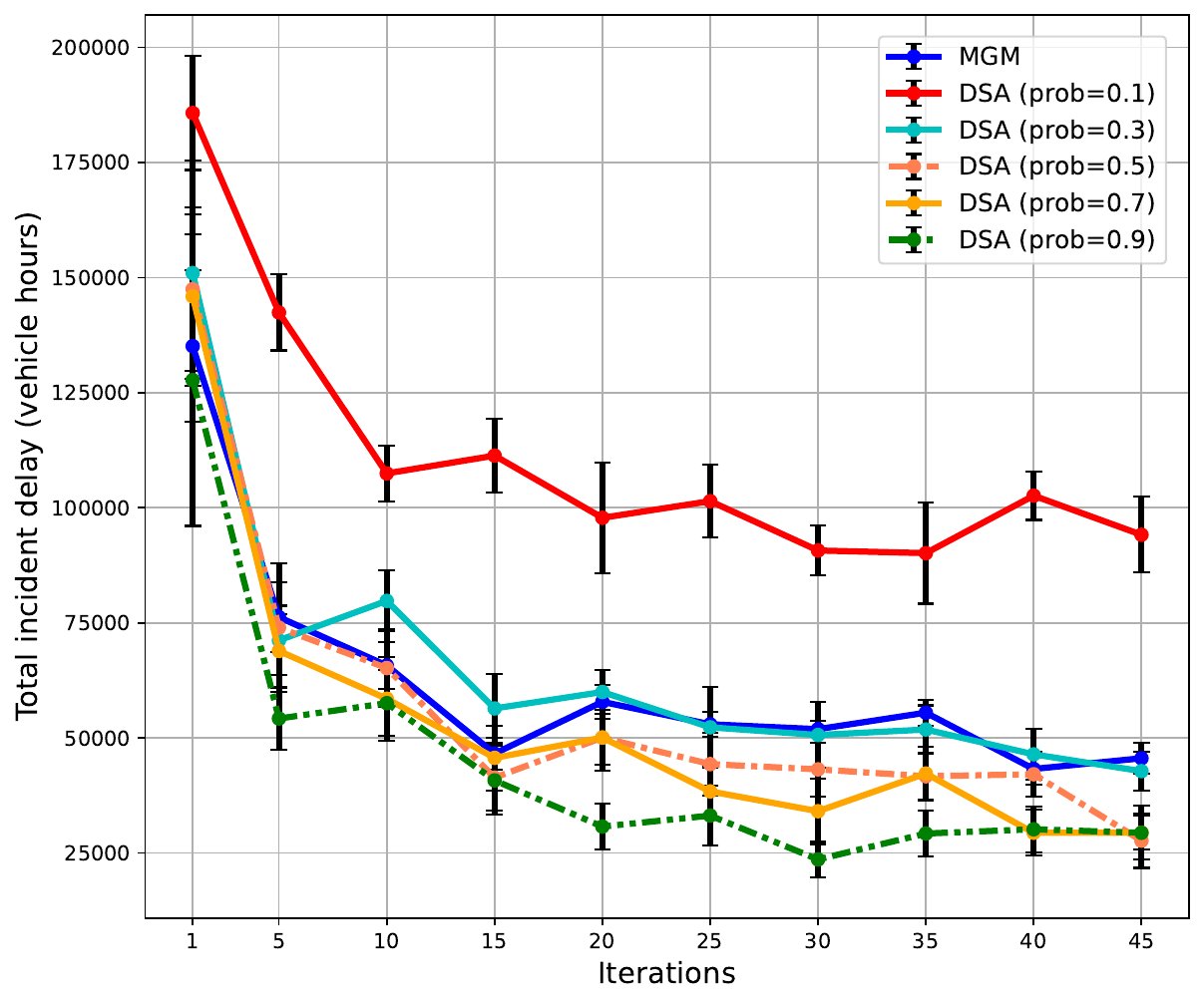}
	\vspace{-2em}
	\caption{MGM and DSA (varying probability threshold) solution and convergence  for total delay impacting network (Incident requests = 5, ERVs = 3.)}
	\label{fig:MGMDSAvprob}
	\vspace{-1.5em}
\end{figure}

Figure \ref{fig:MGMDSAvprob} shows the solution and convergence for MGM and DSA (considering different probability thresholds) for total incident delay for a scenario with five incident requests and three emergency vehicles. The MGM and DSA approaches find better deployment for emergency vehicles as the number of iterations increases. Performing more iterations forces the algorithm to find better solutions than the previous since, in each iteration, the model evaluates other potential solutions. In the DSA approach, the deployment solution improves for higher probability threshold values. The probability threshold determines the level of exploration, and the higher this level, the more likely the algorithm escapes a local optimal to find a better solution. Comparing MGM and DSA, we clearly show the benefit of incorporating exploration over just exploitation. In the case of MGM, the best deployment solution resulted in a total delay of approximately 40000 vehicle hours compared to about 25000 vehicle hours for DSA at a probability threshold of 0.9. Arguably, while MGM got stuck in a local optimal solution, DSA escaped the local optimal through the exploration to find a better deployment solution.

\begin{figure}[!h]
	\centering
	\includegraphics[width=85mm]{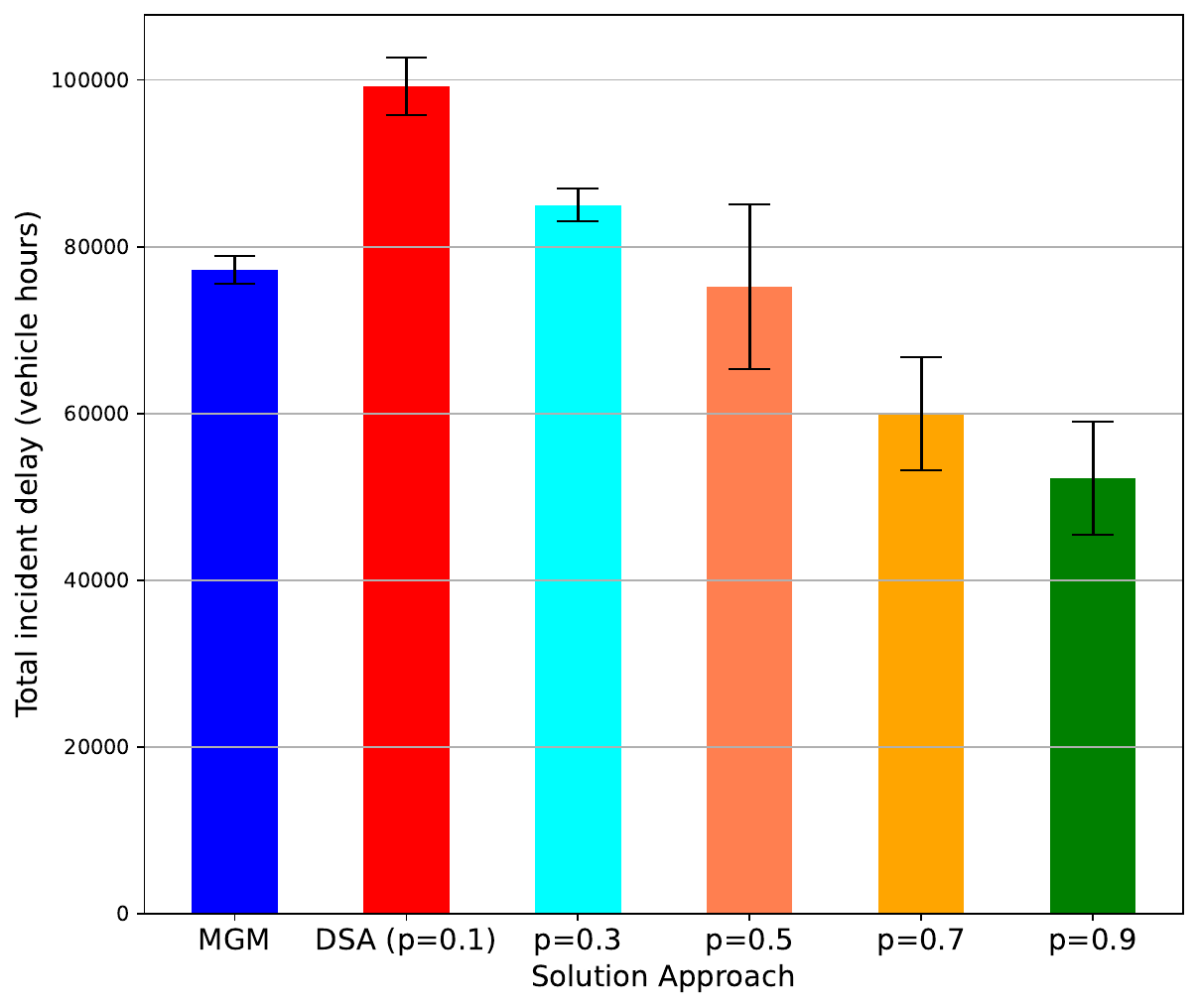}
	\vspace{-2.0em}
	\caption{MGM and DSA solutions for mean incident delay for 100 randomly generated TIM scenarios. }
	\label{fig:csrand}\vspace{-1.0em}
\end{figure}

To further assess the performance of the two solution approach, we conduct a Monte Carlo simulation with 100 scenarios of randomly sampled vehicle numbers, number of incident requests, incident type, and incident locations for MGM and DSA. As seen in Figure \ref{fig:csrand}, implementing the exploration heuristic (DSA) improves the solution for the vehicle assignments, seen with a lower mean total incident delay ($\approx$ 50000 at prob = 0.9) compared to the highly exploitative solution approach (MGM, $\approx$ 78000). Furthermore, DSA finds the best solution at higher probability thresholds. The higher the probability threshold, the more exploration, allowing DSA to find better solutions. Previous studies have confirmed the effect of the probability threshold on the solution quality of DSA.\citep{zhang2005distributed} 

\begin{figure}[!htbp]
	\centering
	\includegraphics[width=85mm]{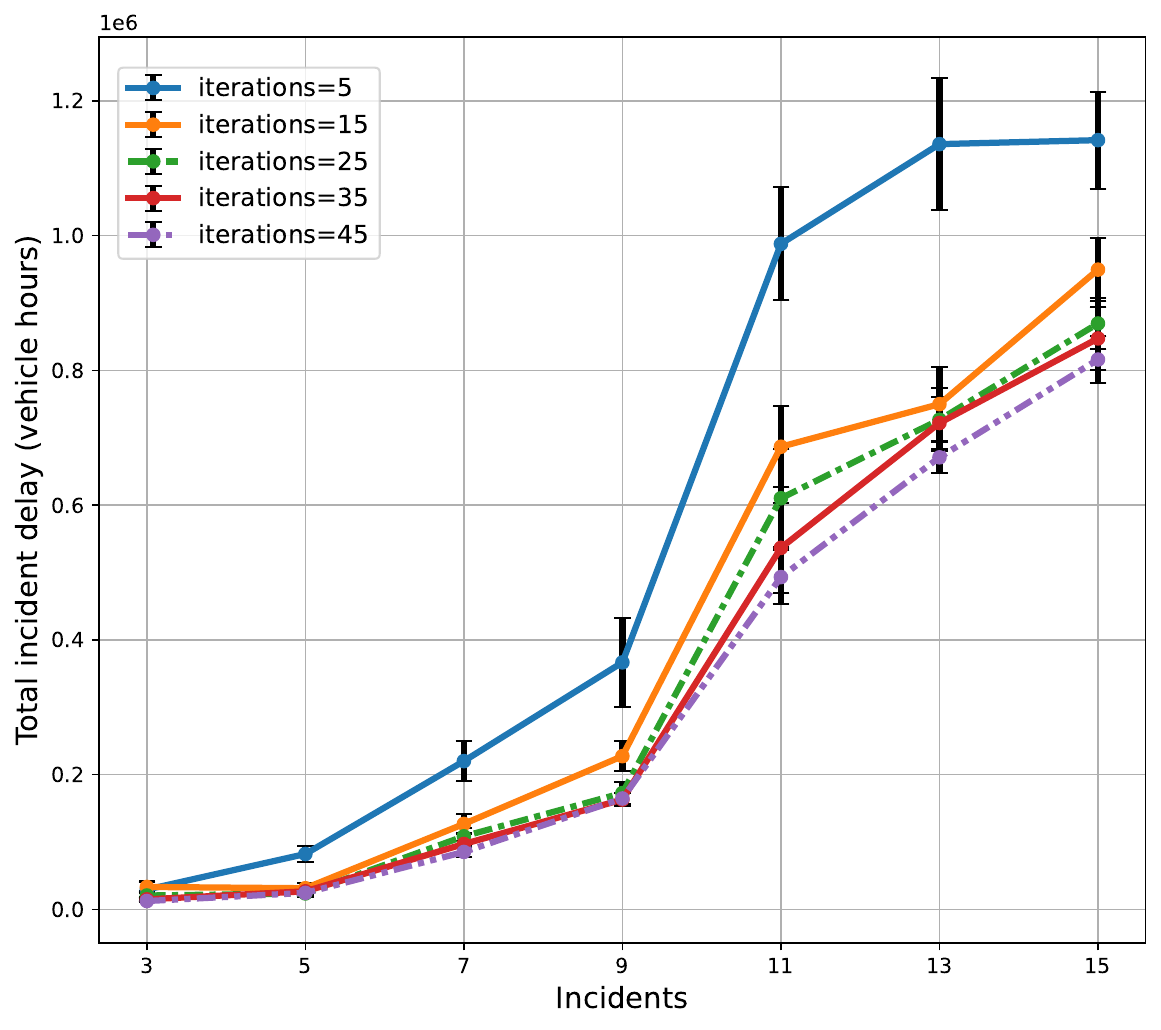}
	\vspace{-2.0em}
	\caption{DSA solution (prob=0.9) for total delay impacting network (ERVs = 3) for different number of incident requests.}
	\label{fig:varincident}
\end{figure} 

By increasing the probability threshold beyond the 0.5 mark, DSA's stochastic exploration technique unveils a remarkable capacity to discover solutions surpassing those of MGM. While this demonstration utilizes an incident request of 5 and an ERVS of 3 in Figure \ref{fig:csrand} and Figure \ref{fig:varincident}, adjusting these parameters can highlight the sensitivity of the probability threshold and flxbility of the DSA preventing the local optimal solution. The subsequent experiments in this comprehensive study leverage the potent DSA approach at a probability threshold of 0.9, providing a rigorous examination of P-DRONETIM's performance across diverse TIM scenarios.

Figure \ref{fig:varincident} illustrates how DSA (with a probability of 0.9) performs for three emergency vehicles under varying numbers of incident requests. As the number of incidents increases, there is a corresponding rise in the total delay that affects the network. Conducting more iterations enhances the ability to search for an optimal deployment solution. For example, performing 45 iterations for fifteen incident requests finds a deployment solution that results in a reduced total incident delay ($\approx$ 800000 vehicle hours) compared to performing only 5 iterations ($\approx$ 1100000 vehicle hours). As the number of incident requests increases from 9 to 11 across all iterations, there is a significant increase in the total incident delay. This is especially remarkable for 5 iterations, but the rate of increase lessens as it approaches 45 iterations.

\color{black}

\begin{figure}[!htbp]
	\centering
	\includegraphics[width=85mm]{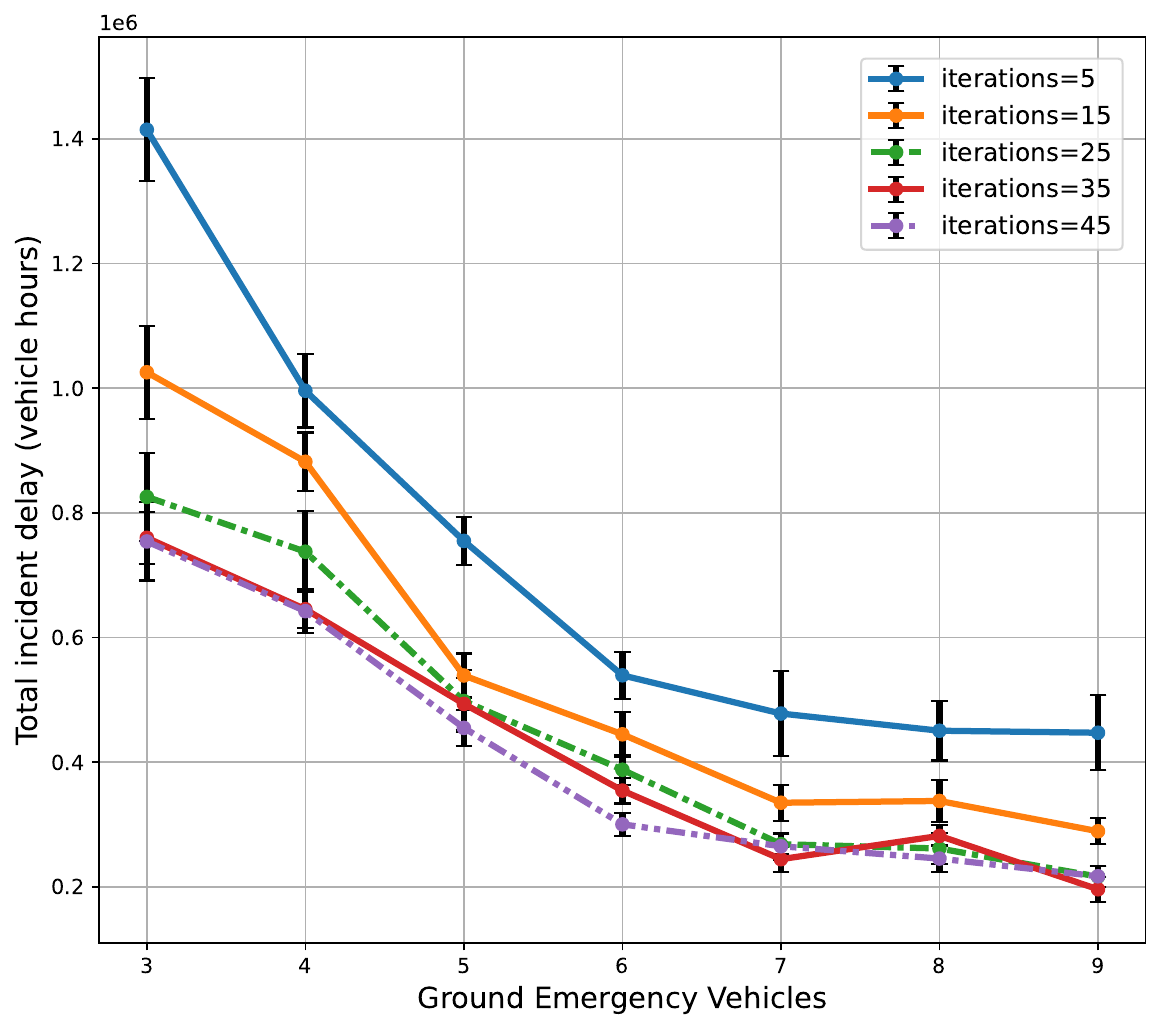}
	\vspace{-2.0em}
	\caption{DSA solution (prob = 0.9) for total delay impacting network for different number of emergency vehicles for scenario with 15 incident requests.}
	\label{fig:DSAvaryvehicle}
\end{figure}

Figure \ref{fig:DSAvaryvehicle} shows the performance of DSA (prob = 0.9) for different number of emergency vehicles for fifteen incident request. Increasing the number of emergency vehicles implies more vehicles available to respond to incident requests and minimize the delay impacting the network. In the case of three emergency vehicles, the vehicles have to respond and clear the initially assigned incidents before it is available to respond to the next incident. The total delay on the network increases when incidents are not responded to quickly. As expected, the higher the number of emergency vehicles, the lower the total delay on the network. For example, considering fifteen incident requests, a team of nine emergency vehicles results in a lower total delay ($\approx 200000$ vehicle hours) than a team with three emergency vehicles ($\approx 750000$ vehicle hours) for 45 iterations. Similar to our prior observation, increasing the number of iterations performed improves the search for the optimal deployment solution.

Similar to the ground emergency vehicles, two experiments are conducted to assess our model's performance for UAV deployment based on the utilities defined by the priority matrix. The first experiment evaluates the effect of varying the number of incidents considering three UAVs. The second experiment evaluates the effect of varying the number of UAVs considering fifteen incident requests.

\begin{figure}[!htbp]
	\centering
	\includegraphics[width=85mm]{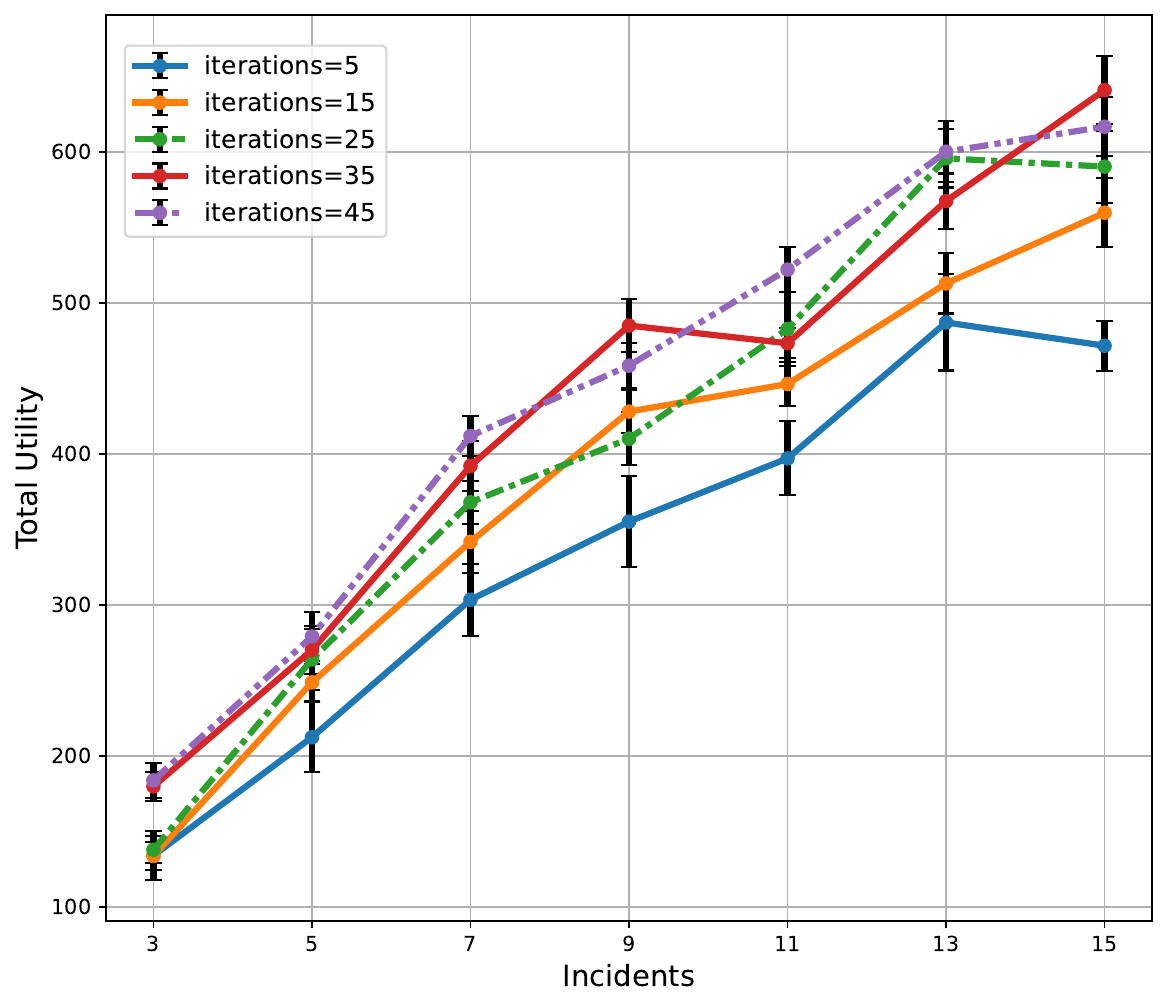}
	\vspace{-2.0em}
	\caption{DSA solution (prob = 0.9) for total utility for different number of incident requests for scenario with 3 UAVs.}
	\label{fig:UAVDSAvaryinc}
\end{figure}

Figure \ref{fig:UAVDSAvaryinc} shows the performance of P-DRONETIM for three UAVs for different number of incident requests. With three UAVs, increasing the number of incidents implies more locations than number of UAVs required for immediate observations. Therefore the UAVs coordinate and decide on a sequence of assignments for each UAV. The results indicate that, as the number of incident requests increases, the UAVs coordinate effectively to provide significant benefits. 

\begin{figure}[!htbp]
	\centering
	\includegraphics[width=85mm]{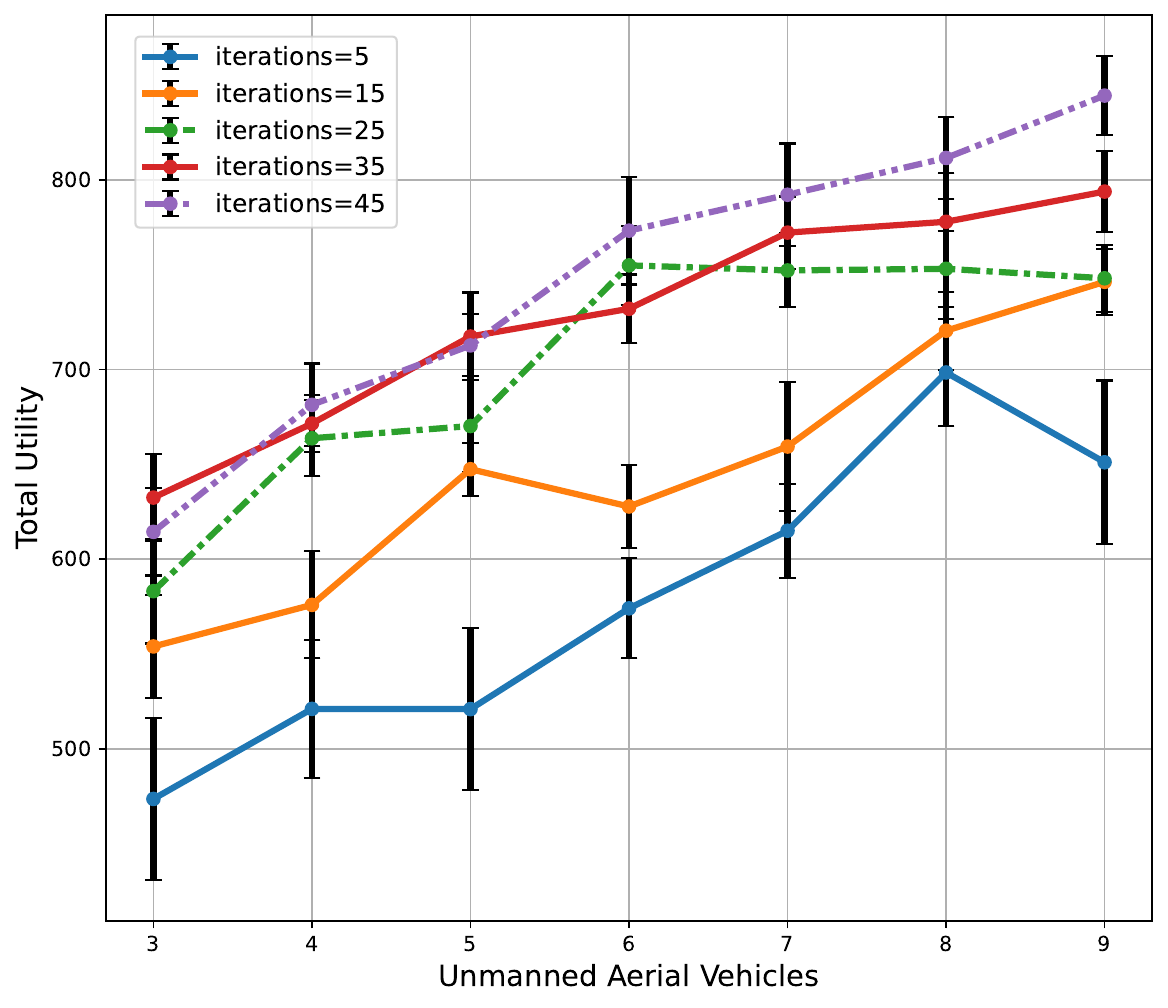}
	\vspace{-2.0em}
	\caption{DSA solution (prob = 0.9) for total utility for different number of UAVs for scenario with 15 incident requests.}
	\label{fig:UAVDSAvaryvehicle}
\end{figure}

Figure \ref{fig:UAVDSAvaryvehicle} shows the performance of P-DRONETIM for different numbers of UAVs considering fifteen incident requests. Increasing the number of UAVs implies more aerial vehicles are available to quickly support the incident tasks and update the estimates of the delay in the network. Therefore, the total benefit for the network increases when more UAVs are available to support the task of ERVs in TIM. For example, considering fifteen incident requests, a fleet of nine UAVs results in a higher total benefit ($\approx 850$) than a fleet with three UAVs ($\approx 610$) for 45 iterations. Increasing the number of iterations performed improves the search for the optimal deployment solution.

\begin{figure}[!htbp]
	\centering
	\includegraphics[width=85mm]{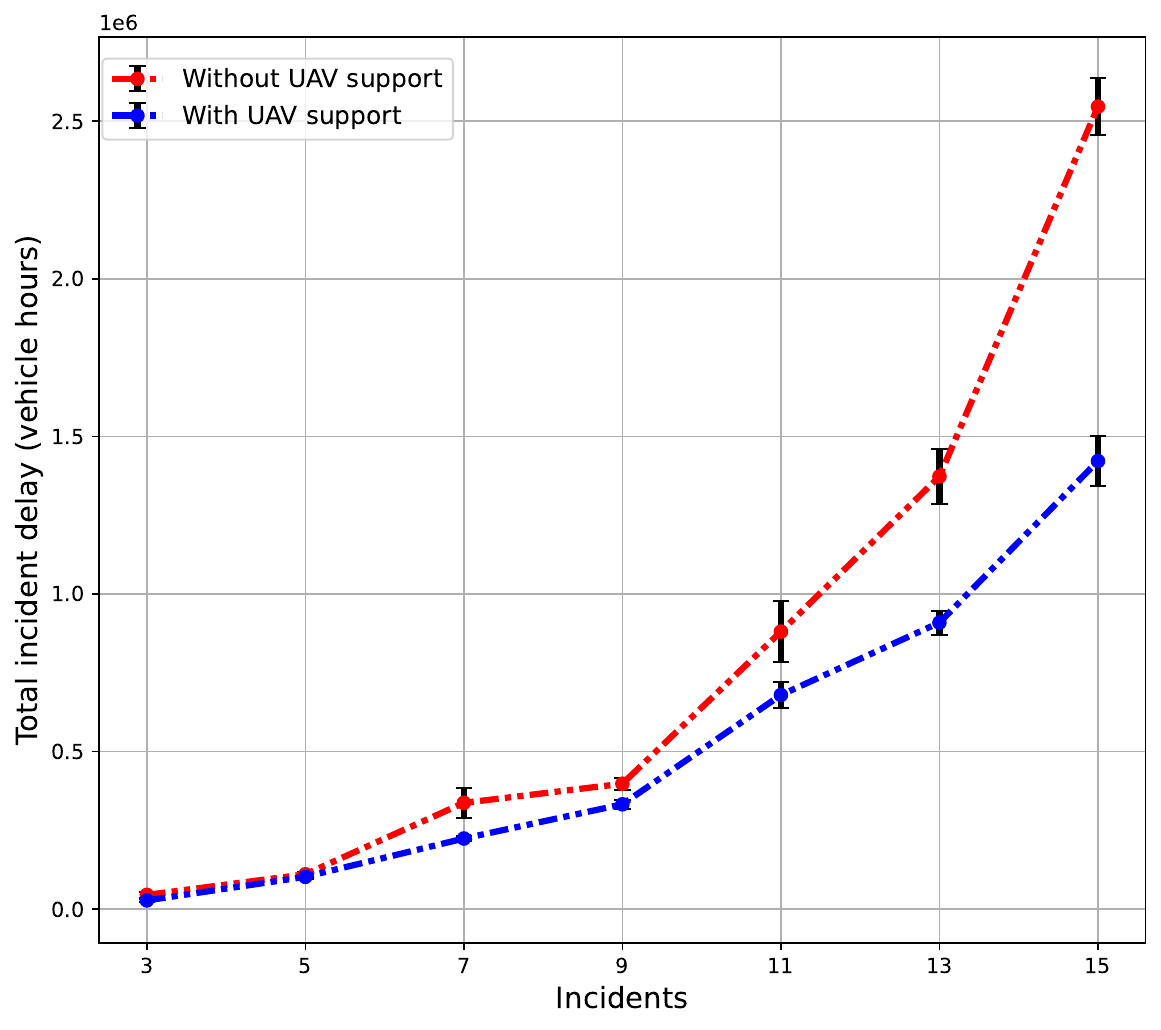}
	\vspace{-2.0em}
	\caption{DSA solution (prob = 0.9) for total incident delay with and without cooperation between ERVs and UAVs.}
	\label{fig:cooperationUAVERV}
\end{figure}

Distributed local search algorithms such as MGM and DSA exhibits favorable time complexity making it suitable for practical implementations. While the algorithm's runtime largely depends on the size of the system and the density of constraints, MGM and DSA typically have fast convergence, an essential property in dynamic environment \citep{zhang2005distributed, maheswaran2004distributed, pearce2007quality}. 

MGMs' ability to efficiently evaluate alternate solutions and select optimal one among them is well documented, even in systems with large constraints density \citep{maheswaran2004distributed, pearce2007quality, yokoo2000algorithms, yokoo1998distributed}. The algorithm's parallelizable nature allows for optimization on modern parallel computing architectures, further enhancing its scalability and applicability to problems of large size \cite{maheswaran2004distributed, pearce2007quality}. 

DSA, as a stochastic algorithm, exhibits variable time complexity, typically converging within a reasonable number of iterations \citep{zhang2005distributed}. DSA algorithms are known for their simplicity and efficiency, requiring fewer iterations compared to other methods. For instance, in contrast to MGM's two cycles per iteration, DSA incurs a cost of one cycle per iteration \citep{zhang2005distributed}. Moreover, DSA's decentralized approach facilitates efficient utilization of computational resources. The algorithm's inherent parallelism allows it to leverage distributed computing environments effectively, making it particularly suitable for systems with large constraints density.

We further perform experiments that assess the effect of cooperation between the two sub-teams of ERVs and UAVs on the total incident delay on the network.
Figure \ref{fig:cooperationUAVERV} shows the total incident delay in the network for the TIM scenario with different numbers of incidents, with and without cooperation, between three ERVs and UAVs. It is clear that when the ERVs have UAV support, there is a lower incident delay compared to when there is no support. This is because the ERVs with UAV support can use the information they receive from the UAVs to improve their response time to incident locations and thus reduce the incident delay impact. Depending on the hazard level on the route, the percentage reduction in response time is estimated to calculate the expected incident duration and delay. 

\begin{figure}[!htbp]
	\centering
	\includegraphics[width=83mm]{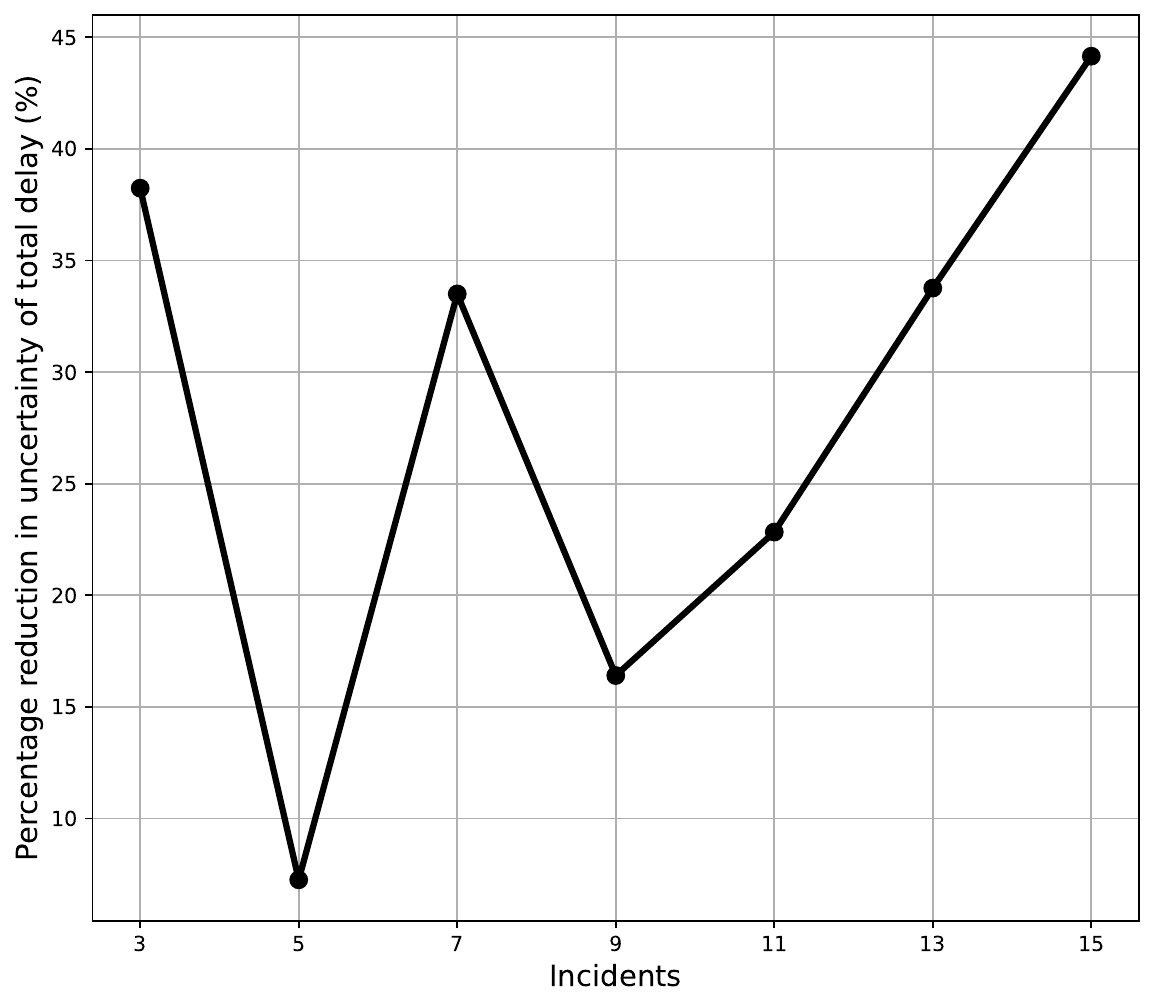}
	\vspace{-1.0em}
	\caption{Percentage reduction in uncertainty of posterior with UAV active sensing of incident location.}
	\label{fig:improvementinposterior}
\end{figure}

Figure \ref{fig:improvementinposterior} assess the level of improvement in incident delay estimation by evaluating the percentage reduction in uncertainty of the total delay when there are observations of incident location by the UAVs. Considering three UAVs, we observe a significant reduction in the estimated total delay ranging between 5 and 45\% for the different number of incidents. UAV actively sensing the incident locations provides enhanced observations used to update the posterior estimates of the expected delay impact as described in Equations \ref{eq:be} and \ref{eq:be1}. 

In general, UAVs face limitations in terms of operation time due to constraints on battery capacity. The onboard circuitry and sensors contribute to energy consumption, thereby restricting the duration of UAV missions. The proposed algorithms, namely the Maximum Gain Method (MGM) and the Distributed Stochastic Algorithm (DSA), are designed with a focus on algorithmic efficiency, balancing algorithmic complexity and energy efficiency, with the goal of achieving high-quality solutions. Although developing specific energy efficient strategies are beyond the current scope of our study, the incomplete characteristics of these algorithms is designed to achieve convergence within a reasonable timeframe, allowing them to quickly and efficiently find good solutions \citep{maheswaran2004distributed, zhang2005distributed}. These algorithms offer substantial potential alignment with UAV operational restrictions, even if they may not fully satisfy UAV energy requirements.

To further mitigate the impact of limited UAV operation time, future algorithms within P-DRONETIM can incorporate adaptive strategies. UAVs can intelligently decide when to activate sensors or perform computations based on the urgency of the task or available energy resources. This adaptability ensures the efficient use of the limited onboard power. This can be similar to energy-aware priority function that assesses the remaining energy in the vehicle at the end of each interval and compares it with the energy required to reach the desired exit state. If the remaining energy exceeds the requirement, the priority function directs the vehicle to continue searching. Conversely, if the remaining energy is insufficient for a direct path to the exit, the function prioritizes an immediate journey to the exit. This adaptive decision-making process ensures efficient energy utilization \citep{gramajo2017efficient, gramajo2016path}.

Depending on the type of UAV (e.g., fixed wing), it becomes possible to design and implement sampling-based path algorithms to generate flight patterns that leverage air flow dynamics to mitigate energy consumption \citep{JDESS}. Such a method aims to improve the UAV's overall coverage distance in accordance with energy-conscious strategies.

Last but not least, building upon the mutually beneficial connections between UAVs and Ground Emergency Vehicles, in future studies, UAV energy constraint in P-DRONETIM can be addressed by refining the algorithm to make dynamic decisions regarding when and where to recharge. This could involve assessing the energy levels of the UAV, considering the proximity to recharge stations, and strategically planning recharging intervals to maximize mission coverage. For instance, in addition to the stationary charging infrastructure, portable recharge stations can be made available on each Ground Emergency Vehicle, further promoting coverage and operational efficiency.

Considering three emergency vehicles, we evaluate the performance of the proposed model against the conventional model in multiple experiments. Each experiment has different sequences of incident requests. We consider five incident request stages for which new incident numbers are reported. 

\begin{figure*}[!t]
	\centering
	\includegraphics[width=170mm]{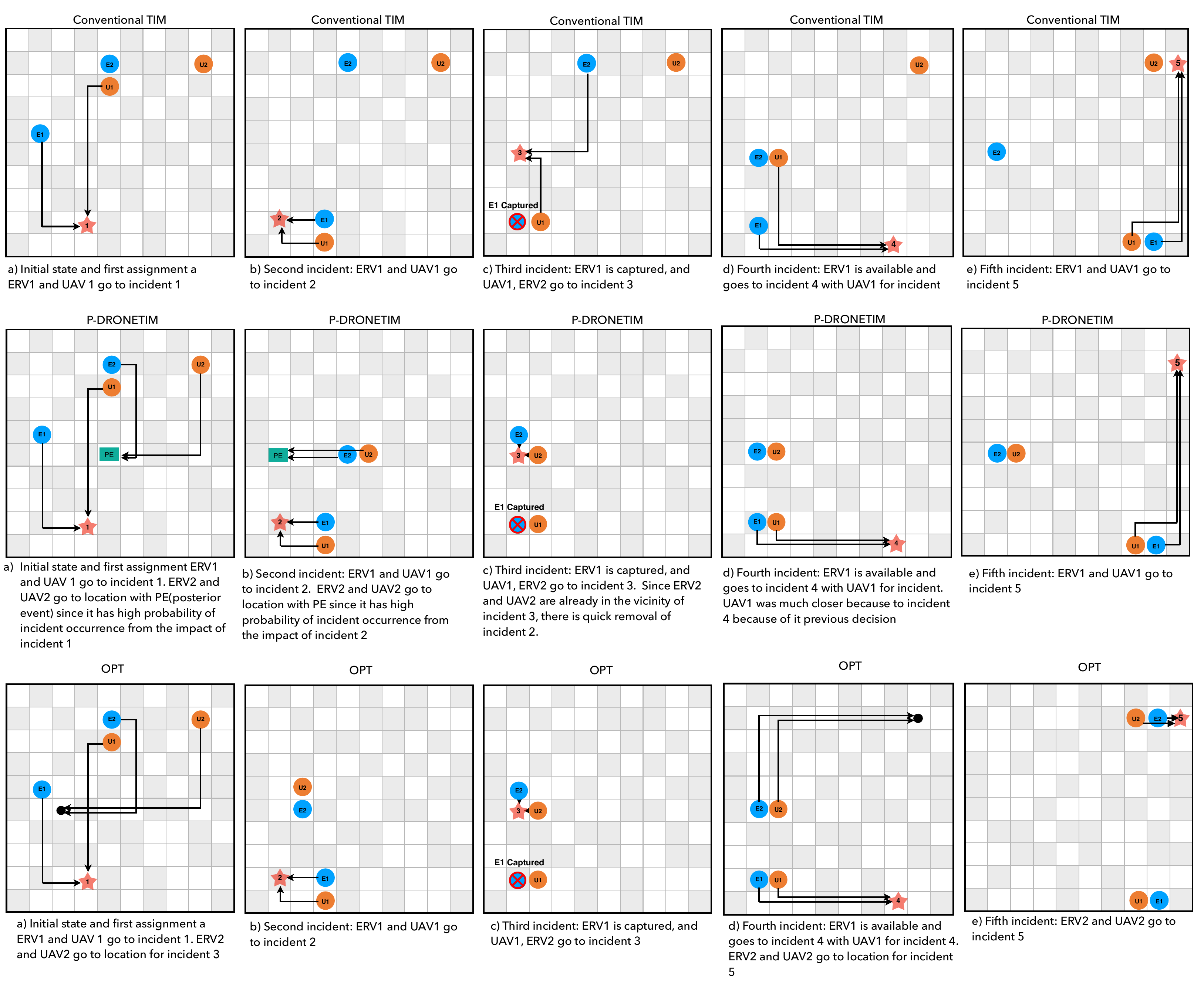}
	\vspace{-1.5em}
	\caption{Runtime example for two ERVs and UAVs for five incidents with different request times.}
	\label{fig:runtimeexample}
\end{figure*}

\begin{table}[!ht]
	\centering
	\caption{Comparison of Total incident Delay (vehicle hours) for Three Models for Different Sequences of Incident Request}
	\vspace{-0.3em}
	\begin{tabular}{|c|c|c|c|c|}
		\hline
		$\text{Scenario}$ & $\text{Conventional TIM}$ &$\text{P-DRONETIM}$ & $\text{OPT}$ \\\hline
		3,2,2,2,1   & 72002  & 68200  & 62200 	 \\\hline
		2,2,2,1,2   & 70330  & 60300  & 58211  \\\hline
		4,2,1,3,3	& 75000  & 63200  & 60235 	 \\\hline
		5,3,4,2,2 	& 88430  & 78500  & 68112 	 \\\hline
		1,3,5,2,1 	& 73500  & 71500  & 65170 	 \\\hline
		2,5,1,1,3 	& 74340  & 70100  & 65981 	 \\\hline
		3,3,3,2,1   & 73800  & 69501  & 63730 	 \\\hline
	\end{tabular}
	\label{Tab:comparison}
\end{table}

For example, in Table \ref{Tab:comparison}, scenario one (3,2,2,2,1) indicates that three incidents were reported in the first stage, followed by two in the next stage, in that order. The total incident delay for the deployment solution for three models is presented. First, the conventional TIM model makes current stage decisions without considering near-future stage decisions. Second, the proposed P-DRONETIM considers the near-future stage decisions when making the current stage decision. Finally, the optimal strategy model (OPT) assumes knowledge of the exact dependencies and sequences of the incident and thus makes the best deployment decision. 

As shown in Table \ref{Tab:comparison}, P-DRONETIM showed significant improvement when compared to the conventional TIM. Specifically, the three emergency vehicles successfully coordinated to find the best sequences of assignments in each current stage. On average, P-DRONETIM reduced the total delay impact by approximately 8\% than the conventional TIM. The minimum and maximum improvements were estimated as 3\% and 15\%, respectively. Compared to OPT, P-DRONETIMs percentage improvement decreased, ranging between 3\% and 13\%, but much worse when OPT is compared to conventional TIM (11\% and 22\%). OPT finds the best deployment solution since it assumes knowledge of the exact sequence of incident requests.

\vspace{-0.1em} Figure \ref{fig:runtimeexample} shows the runtime example for a scenario with five incident requests in sequence (1,1,1,1,1) with no concurrent emergencies at different times. Incidents requests one to four have a severity of 1, and incident request five has a severity of 4. The response time estimates for the conventional TIM model ranges between 3.6 min and 17.4 min, with a total response time of 62.4 min and a total estimated incident delay of 50022 vehicle hours. ERV1 is dispatched to incident requests one, two, four, and five, and ERV2 to incident requests three and four. Similarly, P-DRONETIM dispatched ERV1 to incident requests one, two, four, and five and ERV2 to incident three. 

However, ERV2 was relocated twice based on the estimated probability of incident occurrences at the request times of incidents one and two. With this look-ahead to anticipate the near-future incident, ERV2's response time to incident request three is drastically reduced since it is already in the vicinity of incident request three. The total response time for P-DRONETIM is estimated as 59.1 min, with a total incident delay of 45200 vehicle hours representing a 5.28\% reduction in total response time and 9.64\% reduction in total incident delay compared to the traditional TIM as DCOP strategy. 

In OPT, ERV1 is dispatched to incident requests one, two, and four, and ERV2 dispatches ERV2 to incident requests three and five. Since OPT assumes to know precisely where the incident will happen, it makes the best relocation decisions at the different request times. The total response time and incident delay for OPT is estimated as 51.5 min and 40520 vehicle hours. Compared to OPT, P-DRONETIM's percentage improvement in response and incident delay fell short by 12\% and 10\%, but much worse when OPT is compared to conventional TIM (17\% and 19\%). 

\section{CONCLUSIONS}\vspace{-0.5em}
Traditionally, traffic incident management programs coordinate the deployment of emergency resources to immediate incident requests without anticipating near future requests and the associated interdependencies in the network. In this work, a new proactive and dynamic framework, P-DRONETIM, based on the distributed constraint optimization problem, provides a roadmap for practical resource allocation for highway incident management by accommodating interdependencies in network events. Furthermore, the optimization objective is formulated to accommodate near future stage decisions when making current stage decisions. In addition, we develop a new framework for allocating UAVs and coordinating their benefits for emergency vehicles' response time to incident locations and improving the estimates of the total delay impact.

The proposed framework is solved with the DCOP local search algorithms MGM and DSA. In multiple experiments, P-DRONETIM showed satisfactory performance. P-DRONETIM model actively relocates emergency vehicles to anticipate future incidents to improve the overall response time to incident requests and showed significant performance improvement compared to the conventional TIM. Furthermore, coordinating the benefit of UAV assignment for ERVs resulted in further improvement in performance.

Future research can enhance the allocation process for UAVs by accommodating its routing decisions. For example, the routing decision can be modeled to maximize the reduction in entropy for observed traffic parameters such as link travel time,\citep{LF, JDESS, KDD} shockwave parameters, and discharge rate resulting from an incident while en-route to an assigned location. In addition, to optimize the monitoring of locations by UAVs and manage the limited resources, a coordination constraint that considers deployments having overlapping surface areas can be investigated. Such a constraint will limit the assignment of UAVs to the same neighborhood.

To facilitate the future implementation of aerial resource allocation and support the deployment of Drones as First Responders, the Federal Aviation Administration must approve waivers for Beyond Visual Line of Sight (BVLOS) operations. There are several challenges associated with using human pilots as visual observers for UAVs: 1) the need to keep licensed pilots up to date with regular training; 2) the high expense of employing human pilots; and 3) the requirement for pilots to remain on standby 24/7, leading to considerable costs and posing a significant bottleneck. After the decision on UAV allocation is made, real-world dispatching will need to occur through BVLOS without human pilots, utilizing advanced detect-and-avoid technologies in the airspace, which is currently a subject of emerging research in the field.

\color{black}

\begin{funding}
Funding for this research was provided by the NSF RI \#1910397, NASA JPL RSA \#1625294, and USDOT University Transportation Centers (Contract 69A3551747125). 
\end{funding}

\begin{acks}
This work has been previously shared as a preprint on arXiv under the  title "Proactive Distributed Constraint Optimization of Heterogeneous Incident Vehicle Teams" \citep{darko2022proactive}, and also as part of the PhD dissertation titled "Active Sensing and Learning Multimodality for Predictive Decision Making" \citep{darko2022active}. The authors would also like to express thanks to all editors and reviewers of this article.
\end{acks}

\subsection*{\normalsize\sagesf\bfseries Data Availability}
The data utilized in this study were generated through simulations designed specifically for the presented research. Detailed descriptions of the simulation setup, scenarios, and
experimental parameters are provided in the "Scenarios and Experiment" section of the
article.

\subsection*{\normalsize\sagesf\bfseries Conflict of Interest}
The authors declare that they have no conflicts of interest related to the research,
authorship, or publication of this manuscript.

\bibliographystyle{SageV}

\begin{thebibliography}{99}

\bibitem{fhwa_fsp}
"FHWA FSP Handbook: Chapter 4. Full-Function Service Patrol Concept," U.S. Department of Transportation/Federal Highway Administration, 2008. [Online]. Available: \url{https://ops.fhwa.dot.gov/publications/fhwahop08031/fsp4_0.htm}

\bibitem{daneshgar2013evaluating}
F. Daneshgar, S. P. Mattingly, and A. Haghani, "Evaluating beat structure and truck allocation for the Tarrant County, Texas, courtesy patrol," \textit{Transportation Research Record}, vol. 2334, no. 1, pp. 40--49, 2013.

\bibitem{park2016real}
H. Park and A. Haghani, "Real-time prediction of secondary incident occurrences using vehicle probe data," \textit{Transportation Research Part C: Emerging Technologies}, vol. 70, pp. 69--85, 2016.

\bibitem{pugh2019prediction}
N. Pugh and H. Park, "Prediction of Secondary Crash Likelihood considering Incident Duration using High Order Markov Model," in \textit{IEEE SoutheastCon}, Huntsville, AL, pp. 11--14, 2019.

\bibitem{NEXTGEN_FHWA}
"Next-generation TIM: Integrating Technology, Data, and Training," U.S. Department of Transportation/Federal Highway Administration. [Online]. Available: \url{https://www.fhwa.dot.gov/innovation/everydaycounts/edc_6/nextgen_tim.cfm}

\bibitem{Modi}
P. J. Modi, W. Shen, M. Tambe, and M. Yokoo, "ADOPT: Asynchronous distributed constraint optimization with quality guarantees," \textit{Artificial Intelligence}, vol. 161, no. 1-2, pp. 149--180, 2005.

\bibitem{Yeoh}
W. Yeoh and M. Yokoo, "Distributed problem solving," \textit{AI Magazine}, vol. 33, no. 3, pp. 53--53, 2012.

\bibitem{deqi2012simulation}
H. Deqi, C. Xiumin, and M. Zhe, "A simulation framework for emergency response of highway traffic accident," \textit{Procedia Engineering}, vol. 29, pp. 1075--1080, 2012.

\bibitem{duan2015emergency}
X. Duan, S. Song, J. Zhao, and others, "Emergency vehicle dispatching and redistribution in highway network based on bilevel programming," \textit{Mathematical Problems in Engineering}, vol. 2015, 2015.

\bibitem{Yin}
Y. Yin, "Optimal fleet allocation of freeway service patrols," \textit{Networks and Spatial Economics}, vol. 6, no. 3, pp. 221--234, 2006.

\bibitem{wang2020research}
P. Wang, J. Yang, Y. Jin, and J. Wang, "Research on allocation and dispatching strategies of rescue vehicles in emergency situation on the freeway," in \textit{Proc. 16th International Conference on Control, Automation, Robotics and Vision (ICARCV)}, 2020, pp. 130--135.

\bibitem{Yin1}
Y. Yin, "A scenario-based model for fleet allocation of freeway service patrols," \textit{Networks and Spatial Economics}, vol. 8, no. 4, pp. 407--417, 2008.

\bibitem{Park1}
H. Park, D. Waddell, and A. Haghani, "Online optimization with look-ahead for freeway emergency vehicle dispatching considering availability," \textit{Transportation Research Part C: Emerging Technologies}, vol. 109, pp. 95--116, 2019.

\bibitem{JD2}
D. Justice, L. Folsom, N. Deshpande, and H. Park, "Distributed Constraint Optimization Problem for Coordinated Response of Unmanned Aerial Vehicles and Ground Vehicles," in \textit{Proc. 55th Annual Conference on Information Sciences and Systems (CISS)}, 2021.

\bibitem{FAA_drones}
"Unmanned Aircraft Systems (UAS)," U.S. Department of Transportation/Federal Aviation Administration. [Online]. Available: \url{https://www.faa.gov/uas/}

\bibitem{Leaute}
T. Léauté and B. Faltings, "Distributed constraint optimization under stochastic uncertainty," in \textit{Proc. AAAI Conference on Artificial Intelligence}, vol. 25, no. 1, 2011.

\bibitem{Cheung}
B. Cheung, S. Davey, and D. Gray, "Comparison of the PMHT path planning algorithm with the genetic algorithm for multiple platforms," in \textit{Proc. 13th International Conference on Information Fusion}, 2010, pp. 1--8.

\bibitem{Modi1}
P. J. Modi, "Distributed constraint optimization for multiagent systems," Ph.D. dissertation, University of Southern California, 2003.

\bibitem{silaghi2006nogood}
M. C. Silaghi and M. Yokoo, "Nogood based asynchronous distributed optimization (adopt ng)," in \textit{Proc. Fifth International Joint Conference on Autonomous Agents and Multiagent Systems}, 2006, pp. 1389--1396.

\bibitem{petcu2005dpop}
A. Petcu and B. Faltings, "DPOP: A scalable method for multiagent constraint optimization," in \textit{Proc. IJCAI}, 2005, pp. 266--271.

\bibitem{modi2003asynchronous}
P. J. Modi, W. Shen, M. Tambe, and M. Yokoo, "An asynchronous complete method for distributed constraint optimization," in \textit{Proc. AAMAS}, 2003, pp. 161--168.

\bibitem{MR}
R. T. Maheswaran, J. P. Pearce, and M. Tambe, "A family of graphical-game-based algorithms for distributed constraint optimization problems," in \textit{Coordination of Large-Scale Multiagent Systems}, Springer, 2006, pp. 127--146.

\bibitem{junges2008evaluating}
R. Junges and A. L. C. Bazzan, "Evaluating the performance of DCOP algorithms in a real-world, dynamic problem," in \textit{Proc. 7th International Joint Conference on Autonomous Agents and Multiagent Systems}, vol. 2, 2008, pp. 599--606.

\bibitem{Farinelli}
A. Farinelli, A. Rogers, and N. R. Jennings, "Agent-based decentralised coordination for sensor networks using the max-sum algorithm," \textit{Autonomous Agents and Multi-Agent Systems}, vol. 28, no. 3, pp. 337--380, 2014.

\bibitem{grubshtein2010local}
A. Grubshtein, R. Zivan, T. Grinshpoun, and A. Meisels, "Local search for distributed asymmetric optimization," in \textit{Proc. 9th International Conference on Autonomous Agents and Multiagent Systems}, vol. 1, 2010, pp. 1015--1022.

\bibitem{yin2009local}
Z. Yin, C. Kiekintveld, A. Kumar, and M. Tambe, "Local optimal solutions for DCOP: New criteria, bound, and algorithm," in \textit{Proc. 8th Int. Conf. on Autonomous Agents and Multiagent Systems (AAMAS 2009)}, 2009.

\bibitem{drones1}
A. H. Wheeb, R. Nordin, A. A. Samah, M. H. Alsharif, and M. A. Khan, "Topology-Based Routing Protocols and Mobility Models for Flying Ad Hoc Networks: A Contemporary Review and Future Research Directions," \textit{Drones}, vol. 6, no. 1, Article no. 9, 2022.

\bibitem{drones2}
A. H. Wheeb, R. Nordin, A. A. Samah, and D. Kanellopoulos, "Performance Evaluation of Standard and Modified OLSR Protocols for Uncoordinated UAV Ad-Hoc Networks in Search and Rescue Environments," \textit{Electronics}, vol. 12, no. 6, Article no. 1334, 2023.

\bibitem{drones3}
M. T. Naser and A. H. Wheeb, "Implementation of RWP and Gauss Markov Mobility Model for Multi-UAV Networks in Search and Rescue Environment," \textit{International Journal of Interactive Mobile Technologies (iJIM)}, vol. 16, no. 23, pp. 125--137, Dec. 2022.

\bibitem{Parksensor22}
H. Park, A. Haghani, S. Gao, M. A. Knodler, and S. Samuel, "Anticipatory Dynamic Traffic Sensor Location Problems with Connected Vehicle Technologies," \textit{Transportation Science}, vol. 52, no. 6, pp. 1299--1326, 2018.

\bibitem{P2}
H. Park and A. Haghani, "Stochastic capacity adjustment considering secondary incidents," \textit{IEEE Transactions on Intelligent Transportation Systems}, vol. 17, no. 10, pp. 2843--2853, 2016.

\bibitem{transportation2010trb}
"Highway Capacity Manual," Transportation Research Board, 2010.

\bibitem{li2006estimation}
J. Li, C.-J. Lan, and X. Gu, "Estimation of incident delay and its uncertainty on freeway networks," \textit{Transportation Research Record}, vol. 1959, no. 1, pp. 37--45, 2006.

\bibitem{tang2020statistical}
J. Tang, L. Zheng, C. Han, W. Yin, Y. Zhang, Y. Zou, and H. Huang, "Statistical and machine-learning methods for clearance time prediction of road incidents: A methodology review," \textit{Analytic Methods in Accident Research}, vol. 27, Article no. 100123, 2020.

\bibitem{asch2016data}
M. Asch, M. Bocquet, and M. Nodet, \textit{Data Assimilation: Methods, Algorithms, and Applications}, SIAM, 2016.

\bibitem{american2007manual}
"Manual on Classification of Motor Vehicle Traffic Accidents," ANSI, vol. 16, pp. 1--2007, 2007.

\bibitem{martin2011traffic}
P. T. Martin, P. Chaudhuri, I. Tasic, M. Zlatkovic, and T. Pedersen, "Traffic incident management state of the art review," \textit{Civil and Environmental Engineering}, 2011.

\bibitem{tel}
G. Tel, \textit{Introduction to Distributed Algorithms}, Cambridge University Press, 2000.

\bibitem{zhang2005distributed}
W. Zhang, G. Wang, Z. Xing, and L. Wittenburg, "Distributed stochastic search and distributed breakout: Properties, comparison, and applications to constraint optimization problems in sensor networks," \textit{Artificial Intelligence}, vol. 161, no. 1-2, pp. 55--87, 2005.

\bibitem{maheswaran2004distributed}
R. T. Maheswaran, J. P. Pearce, M. Tambe, and others, "Distributed Algorithms for DCOP: A Graphical-Game-Based Approach," in \textit{PDCS}, pp. 432--439, 2004.

\bibitem{pearce2007quality}
J. P. Pearce and M. Tambe, "Quality Guarantees on k-Optimal Solutions for Distributed Constraint Optimization Problems," in \textit{IJCAI}, pp. 1446--1451, 2007.

\bibitem{yokoo2000algorithms}
M. Yokoo and K. Hirayama, "Algorithms for distributed constraint satisfaction: A review," \textit{Autonomous Agents and Multi-Agent Systems}, vol. 3, pp. 185--207, 2000, Springer.

\bibitem{yokoo1998distributed}
M. Yokoo, E. H. Durfee, T. Ishida, and K. Kuwabara, "The distributed constraint satisfaction problem: Formalization and algorithms," \textit{IEEE Transactions on Knowledge and Data Engineering}, vol. 10, no. 5, pp. 673--685, 1998.

\bibitem{gramajo2017efficient}
G. Gramajo and P. Shankar, "An efficient energy constraint based UAV path planning for search and coverage," \textit{International Journal of Aerospace Engineering}, vol. 2017, Article ID 2017, 2017.

\bibitem{gramajo2016path}
G. Gramajo and P. Shankar, "Path-planning for an unmanned aerial vehicle with energy constraint in a search and coverage mission," in \textit{2016 IEEE Green Energy and Systems Conference (IGSEC)}, pp. 1--6, 2016.

\bibitem{JDESS}
D. Justice, L. Folsom, H. Park, M. Minamide, M. Ono, and H. Su, "A sampling-based path planning algorithm for improving observations in tropical cyclones," \textit{Earth and Space Science Open Archive}, pp. 1--17, 2020.

\bibitem{LF}
L. Folsom, M. Ono, K. Otsu, and H. Park, "Scalable information-theoretic path planning for a rover-helicopter team in uncertain environments," \textit{International Journal of Advanced Robotic Systems}, vol. 18, no. 2, Article no. 1729881421999587, 2021.

\bibitem{KDD}
H. Park, D. Justice, N. Deshpande, V. Pandey, H. Su, M. Ono, D. Barkley, L. Folsom, D. Posselt, and S. Chien, "Temporal Multimodal Multivariate Learning," in \textit{Proc. 28th ACM SIGKDD Conference on Knowledge Discovery and Data Mining}, Washington, DC, pp. 3722--3732, 2022.

\bibitem{darko2022proactive}
J. Darko and H. Park, "Proactive Distributed Constraint Optimization of Heterogeneous Incident Vehicle Teams," \textit{arXiv preprint arXiv:2207.11132}, 2022.

\bibitem{darko2022active}
J. Darko, \textit{Active Sensing and Learning Multimodality for Predictive Decision Making}, Ph.D. dissertation, North Carolina Agricultural and Technical State University, 2022.


\end{thebibliography}

\end{document}